\pdfoutput=1
\documentclass[journal]{IEEEtran}
\usepackage{blindtext}
\usepackage{graphicx}
\usepackage[T1]{fontenc}
\usepackage{textcomp}
\usepackage{lmodern}
\usepackage[latin1]{inputenc}
\usepackage[swedish,english]{babel}
\usepackage{graphics}
\usepackage{amssymb}
\usepackage{amsmath}
\usepackage[hyphens]{url}
\usepackage[hidelinks]{hyperref}
\hypersetup{breaklinks=true}
\usepackage[flushleft]{threeparttable}
\usepackage{stfloats}
\makeatletter
\adddialect\l@ENGLISH\l@english
\makeatother

\ifCLASSINFOpdf
\else
\fi

\newcommand{\MYfooter}{\smash{\scriptsize
\hfil\parbox[t][\height][t]{\textwidth}{\centering
~\\
~}\hfil\hbox{}}}

\newcommand{\MYarxivheader}{\smash{\scriptsize
\hfil\parbox[t][\height][t]{\textwidth}{\centering
This work has been submitted to the IEEE for possible publication. Copyright may be transferred without notice, after which this version may no longer be available.
}\hfil\hbox{}}}

\makeatletter

\def\ps@headings{%
\def\@oddhead{\mbox{}\scriptsize\rightmark \hfil \thepage}
\def\@evenhead{\scriptsize\thepage \hfil \leftmark\mbox{}}
\def\@oddfoot{\MYfooter}%
\def\@evenfoot{\MYfooter}}

\def\ps@IEEEtitlepagestyle{%
\def\@oddhead{\MYarxivheader}%
\def\@evenhead{\scriptsize\thepage \hfil \leftmark\mbox{}}%
\def\@oddfoot{\MYfooter}%
\def\@evenfoot{\MYfooter}}

\makeatother

\begin{document}

% Command to turn off the "------" author when two succeeding entries have the same authors.
%\bstctlcite{IEEEexample:BSTcontrol}

\title{Spin-Torque and Spin-Hall Nano-Oscillators}

\author{Tingsu Chen,~\IEEEmembership{Student Member,~IEEE,}
        Randy K. Dumas,~\IEEEmembership{Member,~IEEE,}
        Anders Eklund,~\IEEEmembership{Student Member,~IEEE,}
        Pranaba K. Muduli,~\IEEEmembership{Member,~IEEE,}
        Afshin Houshang,~\IEEEmembership{Student Member,~IEEE,}
        Ahmad A. Awad,~\IEEEmembership{Member,~IEEE,}
        Philipp D\"urrenfeld,~\IEEEmembership{Student Member,~IEEE,}
        B. Gunnar Malm,~\IEEEmembership{Senior Member,~IEEE,}
        Ana Rusu,~\IEEEmembership{Member,~IEEE,}
        and~Johan~\AA kerman,~\IEEEmembership{Member,~IEEE}% <-this % stops a space
\thanks{T. Chen, A. Eklund, B. G. Malm, and A. Rusu are with the Department of Integrated Devices and Circuits, School of Information and Communication Technology, KTH Royal Institute of Technology, 164 40 Kista, Sweden (e-mail: tingsu@kth.se, ajeklund@kth.se, gunta@kth.se and arusu@kth.se).}
\thanks{R. K. Dumas, P. K. Muduli, A. Houshang, A. A. Awad,  P. D\"urrenfeld and J. $\AA$kerman are with the Department of Physics, University of Gothenburg, Gothenburg, 416 96 Sweden (e-mail: randydumas@gmail.com, pranaba.muduli@physics.gu.se, afshin.houshang@physics.gu.se, ahmad.awad@physics.gu.se, philipp.durrenfeld@physics.gu.se).}% <-this % stops a space
\thanks{P. K. Muduli is also with the Department of Physics, Indian Institute of Technology, Delhi, HauzKhas, New Delhi, 110016, India}
\thanks{J. $\AA$kerman is also with the Department of Materials and Nano Physics, KTH Royal Institute of Technology, 164 40 Kista, Sweden, and Nanosc AB, 164 40 Kista, Sweden. (e-mail: akerman1@kth.se).}% <-this % stops a space
\thanks{Manuscript submitted 8 Oct. 2015}}

%\markboth{Journal of \LaTeX\ Class Files,~Vol.~6, No.~1, January~2007}%
\markboth{Manuscript in preparation for Proceedings of the IEEE}
{Chen \MakeLowercase{\textit{et al.}}: STNOs and SHNOs}


\IEEEspecialpapernotice{(Invited Paper)}

\maketitle

\begin{abstract}

This paper reviews the state of the art in spin-torque and spin Hall effect driven nano-oscillators. After a brief introduction to the underlying physics, %and the applicable oscillator models, 
the authors discuss different implementations of these oscillators, their functional properties in terms of frequency range, output power, phase noise, and modulation rates, and their inherent propensity for mutual synchronization. Finally, the %future 
potential for these oscillators in a wide range of applications, from microwave signal sources and detectors to neuromorphic computation elements, is discussed together with the specific electronic circuitry that has so far been designed to harness this potential. 

\end{abstract}

\begin{IEEEkeywords}
Spintronics, Microwaves, Spin transfer torque, Spin Hall effect.
\end{IEEEkeywords}

% \fi
\IEEEpeerreviewmaketitle

\section{Introduction}
This year marks the $20^{th}$ anniversary of %the publication of the 
two seminal papers by Slonczewski \cite{Slonczewski1996a} and Berger \cite{Berger1996} on the spin transfer torque (STT) phenomenon. This refers to %\emph{i.e.} 
the transfer of angular momentum from a spin-polarized charge current, or a pure spin current, to a local magnetization. While Berger had already reported on STT-driven domain wall motion using 45 A of current in millimeter-wide films in 1985 \cite{Freitas1985}, based on his predictions from the late 1970s \cite{Berger1978, Berger1979}, and Slonczewski had developed the theory for spin-current-driven exchange coupling in magnetic tunnel junctions in 1989 \cite{Slonczewski1989}, the two papers in 1996 %independently 
predicted that a current flowing perpendicularly to the plane of a magnetic metal multilayer should be able to reorient the magnetization direction in one of the layers. These papers were %also 
published at a time when  magnetic thin film technology had just become capable of beginning to implement these concepts in earnest, with giant magnetoresistance (GMR) \cite{Baibich1988,Binasch1989} multilayers being prepared for hard drive read-head production and the first high-quality magnetic tunnel junctions having %just 
been reported the year before \cite{Miyazaki1995,Moodera1995}. It nevertheless took  a few more years before the first experimental demonstration of STT-driven magnetization precession was reported \cite{Tsoi1998}, followed by reports of STT-driven magnetic switching \cite{Myers1999,Katine2000}. 

In 1999, Hirsch revisited \cite{Hirsch1999} a different phenomenon, the so-called spin Hall effect (SHE), first described by D'yakonov and Perel' in 1971 \cite{Dyakonov1971jetp,Dyakonov1971pl}. The SHE can produce a pure spin current in a direction perpendicular to a charge current \cite{Kato2004,Wunderlich2005}, which can in turn  exert substantial STT on an adjacent magnetic layer. It is thus possible to use the SHE in a nonmagnetic metal to achieve, in principle, all the STT-related functions---such as magnetic switching \cite{Miron2011,Liu2012} and driven magnetization precession \cite{Liu2012prl,demidov2012ntm}---that previously required a separate magnetic layer to spin-polarize the charge current. 

%The most important fundamental consequence of STT is the possibility to manipulate the state of a magnet using direct currents instead of through the  magnetic field generated by the current. 
From a technological standpoint, the two most important consequences of STT are \emph{i}) % on the one hand, 
the possibility of programming a magnetic memory, such as magnetoresistive random-access memory (MRAM) \cite{Akerman2005sci,Engel2005}, using a small direct current through the bit, rather than larger currents through adjacent programming lines, and \emph{ii}) %on the other, 
the possibility of creating nanoscopic and ultratunable microwave signal generators without the need for any semiconductor materials. Our paper deals with the latter technology and reviews the fundamentals and  state of the art of so-called spin transfer torque and spin Hall effect nano-oscillators (STNOs and SHNOs). For more in-depth reviews of STT-based MRAM %spin torque magnetoresistive random access memory (ST-MRAM) 
and STNOs, see \cite{Katine2008,Silva2008,Kim2012,Wang2013, Locatelli2013b} and \cite{Slavin2009a}.

\section{STT and SHE fundamentals}

\subsection{Spin currents}

%It is often said that 
The term \emph{spintronics} is often used to refer to electronics in which  both the charge and the spin of the carriers in a metal or semiconductor are utilized (for reviews of spintronics, see \cite{Zutic2004,Chappert2007}). However, this excludes the very active research currently being undertaken in magnetic insulators; these cannot transfer any charge current, but are excellent conductors of spin currents \cite{Stamps2014,Chumak2015}. A broader, more general and, we would argue, better definition of modern spintronics is any device and technology in which \emph{spin currents} \cite{Huan2014}, with or without additional charge or heat currents, are directly employed and utilized for function. 

As spin is a vector in three-dimensional spin space, and the flow of the spin carriers takes place in three-dimensional real space, the resulting spin current is a $3\times3$ tensor (for illustrative purposes, the spin current may instead be pictured as a flow, in real space, of tiny magnetic moments, also in real space). The tensorial nature of the spin current---where the direction of the flowing spins is, in the general case, independent of the real-space direction of the spin current---provide for very rich physics and a tremendous potential for new device functionality.

In the original formulation of STT, the spin current was assumed to be carried by a spin-polarized charge current, and a number of simplifying assumptions (e.g., device geometry and uniformity of  magnetization) were made such that the flow direction of the charge and spin currents were identical. Most analysis and modelling of STNOs still relies on such assumptions, which tend to be highly accurate in most cases. In this case, the spin-polarized current (and hence the spin current) arises directly from the spin polarization of the electrons at the Fermi surface of the ferromagnetic metal. When an unpolarized current, say in Cu, enters a ferromagnetic layer, it becomes spin-polarized in the so-called spin accumulation zone, which extends both into the Cu and the ferromagnet in proportion to the spin diffusion lengths in each material \cite{Valet1993}. As a result, the current acquires spin polarization well inside the Cu, reaches about half its full spin polarization at the NM/FM interface, and then, over a few nanometers in the FM, rapidly reaches the spin polarization of the bulk of the ferromagnet. The same processes govern how the spin polarization of the current decays as the current leaves the ferromagnet and again enters a nonmagnetic metal. As the spin diffusion length is short in ferromagnets, but can be several hundred nanometers even at room temperature in most ordinary metals, thin magnetic films are highly effective spin filters and the resulting spin-polarized currents can act over quite long distances.

\begin{figure}[tb]
  \begin{center}
  \includegraphics[width=8.4cm]{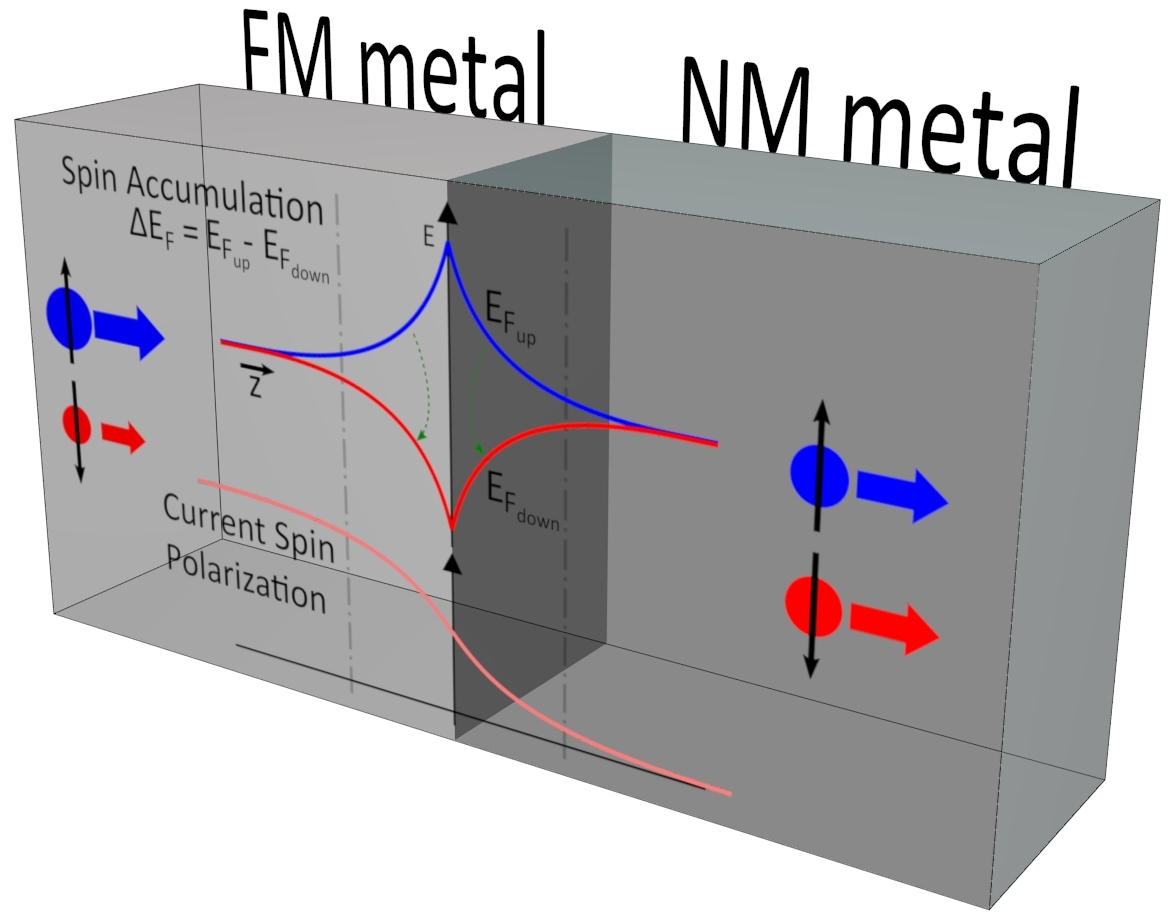}
    \caption{Basic schematic of how spin accumulation zones in both the ferromagnetic (FM) and nonmagnetic (NM) layers arise from transport of a charge current through an FM/NM bilayer. The extent of the zone in each layer depends directly on the spin diffusion length in each layer, which is typically 1--3 orders of magnitude longer in the NM layer than in the FM layer. A very thin FM layer can thus polarize the current over large distances in an NM layer. }
    \label{fig-SpinAccumulation}
    \end{center}
\end{figure}

\subsection{Spin currents generated by the SHE}

While spin filtering is a straightforward technique for generating spin-polarized currents, spin currents do not necessarily require any charge current to flow. In the case of the SHE, %for example, 
the charge and spin currents flow in perpendicular directions with respect to each other---that is, the charge current is unpolarized and the spin current is free of net charge. While we do not discuss this further here,  pure spin currents can also be created through magnetodynamic means, as in so-called spin batteries \cite{Brataas2002}.

The SHE can be viewed as a limiting case of the anomalous Hall effect (AHE) \cite{Hall1881,Nagaosa2010}, which was proposed by Edwin Hall in 1881 when he applied his earlier discovery of the ordinary Hall effect (HE) \cite{Hall1879} to ferromagnetic metals. As is well known, the HE in nonmagnetic metals or semiconductors refers to the build-up of a so-called Hall voltage across an electrical conductor, transverse both to the electric current in the conductor and to an external magnetic field perpendicular to the current. It is widely used for material characterization and for sensing magnetic fields. In ferromagnets, however, Hall found that this effect could be ten times greater than in nonmagnetic metals. 

Whereas the HE arises from the magnetic field's Lorentz force on the charge carriers, the AHE is a consequence of asymmetric spin-dependent scattering of the charge carriers, such that a much larger Hall voltage can build up. The SHE is then a limiting case of this phenomenon in materials without any spontaneous magnetization but with substantial spin-orbit coupling. As there is no spin imbalance in a nonmagnetic material, the asymmetric spin-dependent scattering does not result in any Hall voltage. What it does result in, however, is a net spin current transverse to the charge current and, in a steady state, the build-up of spin accumulation zones at the edges of the conductor, which can be directly observed by optical means \cite{Kato2004,Wunderlich2005} and which is used in spin-current-driven devices.

\subsection{Spin currents generating STT}

Equipped with two different techniques for generating spin currents, we now briefly explain how spin currents can generate STT and control the magnetic states and the magnetodynamics of nanomagnets and magnetic thin films (for an in-depth review of STT, see \cite{Ralph2008a,Sun2008}). 

Since each electron carries angular momentum, a spin current can %therefore 
be thought of as a flow of angular momentum. When a spin-polarized current enters (or reflects at) a nonmagnetic--ferromagnetic interface, the spin precession becomes incoherent over a very short distance.  This results in a loss in its average spin angular momentum in the direction \emph{transverse to the local magnetization}. Since the total angular momentum has to be conserved in a closed system, whenever the flow of angular momentum changes, a torque acts on the local magnetization.  This torque is often referred to as the ``in-plane'' torque, where the plane is defined by the local magnetization and incoming spins. This torque may act gradually, as when spin-polarized electrons flow through a domain wall and have to rotate so as to follow the local magnetization, or very abruptly, as when a spin current enters a ferromagnet with its magnetization in a  direction different to that of the spins in the spin current; this is almost always the case in STNOs and SHNOs.  Such an in-plane torque manifests itself as an energy-nonconserving force, similar to damping.  Furthermore, depending on the sign of the current, this in-plane torque is either parallel or antiparallel to the damping torque. When antiparallel to the damping torque, and for sufficient current magnitude, the current acts to destabilize the local magnetization, thus exciting large angle dynamics. There is also the possibility of an additional STT term that is oriented perpendicular to the plane defined by the local magnetization.  This ``out-of-plane'' torque is often referred to as an ``effective field'' contribution, as it has the same symmetry as is typically used in field-driven dynamics.  While the perpendicular contribution is small for all-metallic multilayers \cite{Stiles2002, Xia2002, Zwierzycki2005}, it can be substantial in devices with tunneling barriers \cite{Xia2002, Theodonis2006, Matsumoto2011}.

\section{STNO and SHNO geometries}

\subsection{STNO Device Architectures}

It was not until sufficient advances in nanofabrication techniques had been achieved that magnetic devices with nanoscopic dimensions could be readily fabricated and the currents needed to induce magnetodynamics be decreased to the mA range.  One of the first techniques relied on mechanical point-contacts, where a highly sharpened metallic tip is used to make electrical contact, typically with an area on the order of 10$^2$ nm$^2$, to an extended multilayer film \cite{Tsoi1998, Tsoi2000, Sun1999}.  Soon after, lithographically defined contacts were employed in devices that are now termed nanocontacts, to distinguish them from their mechanical point-contact predecessors \cite{Myers1999, Rippard2003, Pufall2003, Rippard2004, Dumas2014}.  The fabrication of nanocontacts involves  opening a small electrical contact of either circular or elliptical \cite{Iacocca2015} cross-section through an insulating layer using either electron-beam lithography (EBL) or hole-colloid lithography (HCL) \cite{Sani2013a} techniques.  

In both point-contact and nanocontact devices, it is only the current injection site that is confined to a nanoscopic region.  However, in oscillators based on the nanopillar \cite{Katine2000, Kiselev2003} geometry, all of the magnetic layers are patterned into a wire with either circular or elliptical cross-section.  Fabrication is most often accomplished by a combination of EBL and ion milling.  However, stencil \cite{Sun2002, Ozyilmaz2004, Ozyilmaz2006} and electrodeposition \cite{Wegrowe2002, Araujo2012, Araujo2013} techniques have also been pursued. The nanopillar geometry ensures that all of the electric current passes through the entire multilayer stack, typically resulting in lower current densities (<10$^7$ A/cm$^2$) being needed to excite STT-induced magnetization switching or oscillations than in the case of nanocontact-based devices (10$^8$--10$^9$ A/cm$^2$).  It should be noted that nanopillar fabrication is typically more involved and prone to deleterious edge effects that can increase oscillator linewidths and magnify sample-to-sample variations. Due to the smaller magnetic volumes involved, nanopillar-based devices are also more sensitive to thermally induced fluctuations, which results in larger linewidths than in nanocontact devices \cite{Rippard2006}. Another important consideration is that, as all of the magnetic layers are patterned, significant stray fields are present in nanopillars and need to be considered in analyzing them.  Additionally, hybrid device architectures have also been pursued in which a nanopillar is fabricated on top of an extended film \cite{Demidov2010, Evarts2013}.  Such hybrid structures are useful if, for example, optical access to the extended magnetic layer is desired.   

Generally speaking, the transport of spin current, and therefore of the associated STT, does not need to accompany a charge current.  Such nonlocal devices are engineered to promote different paths for the spin and charge currents.  This geometry is particularly useful in tunnel-junction based devices where large voltages can cause electrical breakdown of the tunnel barrier.  Such devices typically require at least three terminals for both spin current generation and electrical detection. Both STT-induced nonlocal switching \cite{Kimura2006, Yang2008, Sun2009} and oscillations \cite{Xue2012, Demidov2015a} of the magnetization have been demonstrated.  Devices where the STT is provided by the SHE---that is, where the charge and pure spin currents naturally flow in orthogonal directions, are prime examples of nonlocal oscillators \cite{Liu2012prl, Demidov2012, Demidov2014, Duan2014}.   

\begin{figure*}[tb]
  \begin{center}
    \includegraphics[ width=17cm]{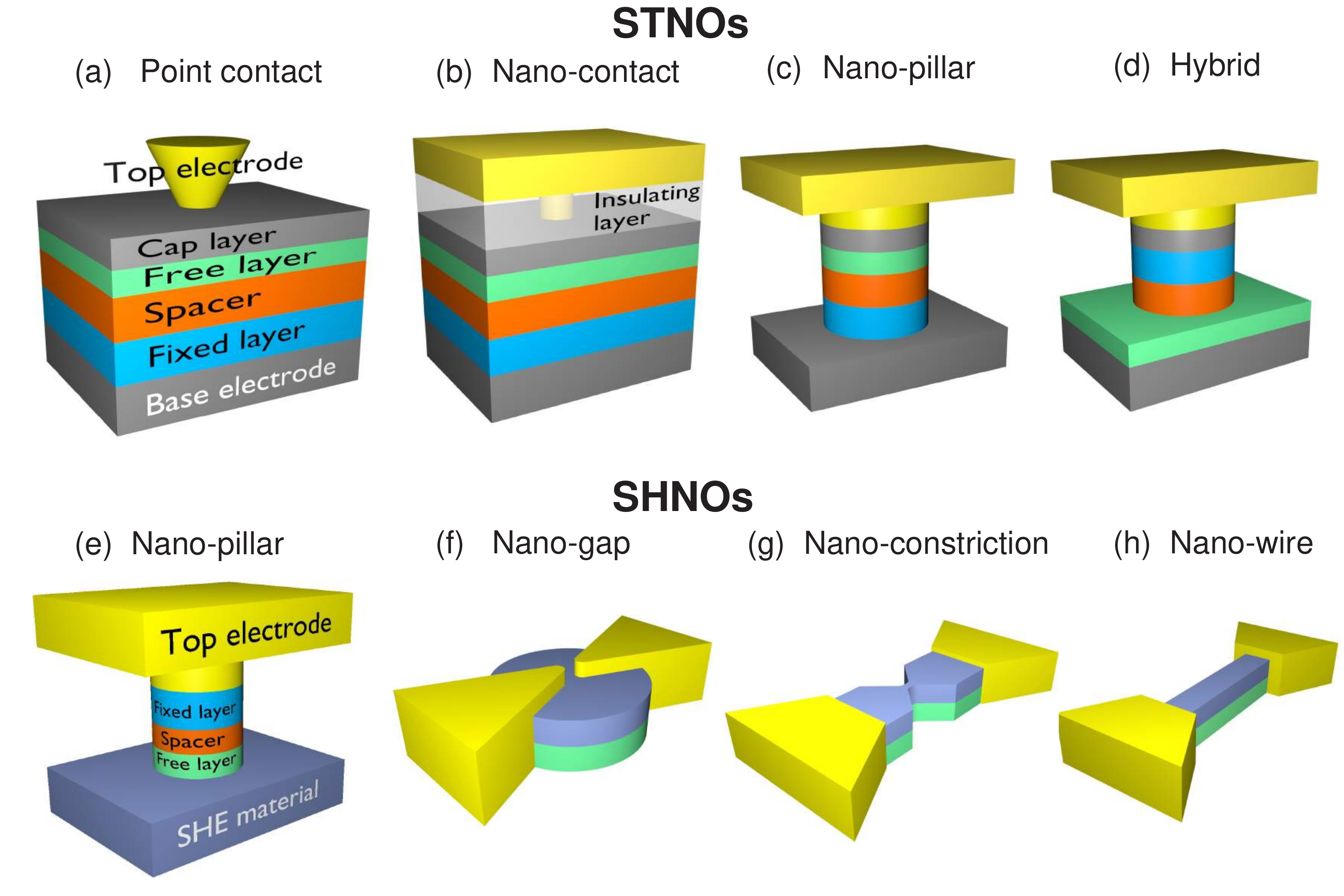}
    \caption{Device schematics}
    \label{fig-mod1}
  \end{center}
\end{figure*}

\subsection{STNO Materials} 

By and large, most material stacks utilized in STNO devices are based on a ferromagnet/spacer/ferromagnet (FM/S/FM) trilayer.  One FM is considered free and is most susceptible to the influence of STT.  It is therefore advantageous for the free layer to be relatively thin and to have a low intrinsic Gilbert damping.  Ni$_{81}$Fe$_{19}$, which goes by the trade name ``Permalloy'' and is often referred to simply as NiFe, is the canonical choice for the free layer in the majority of STNOs and SHNOs.  In contrast, the other FM layer is fixed, either through pinning to an adjacent magnetic layer or by virtue of its intrinsically larger anisotropy or thickness. The primary role of the fixed layer is to spin-polarize the initially unpolarized electron current.  The spacer layer may be metallic (most typically Cu),  forming what is referred to as a pseudo spin valve (PSV), or insulating, as in the case of magnetic tunnel junctions (MTJs).  As the tunneling magnetoresistance (TMR) effect is much larger than the giant magnetoresistance (GMR) effect in all-metallic PSVs, MTJs are becoming standard in most nanopillar devices \cite{Deac2008}.  The use of MTJs in conventional nanocontacts is challenging, as the topmost FM layer would act to shunt the current.  Nevertheless, there has been some success in creating hybrid nanocontact MTJs, often termed ``sombrero'' structures, which aim to marry the low linewidths inherent in nanocontact devices with the large output powers found in MTJs \cite{Maehara2013, Maehara2014}.  

The materials that constitute the FM layers are diverse and can be broadly classified and distinguished by their equilibrium magnetization configurations, as described next.   

\subsubsection{All in-plane}

The vast majority of STNO devices studied in the early stages were based on free and fixed FM layers with an intrinsic easy-plane anisotropy---meaning that the magnetization prefers to lie in the plane of the film if no external magnetic field is applied %in-plane shape anisotropy 
\cite{Rippard2003, Pufall2003, Rippard2004, Katine2000, Kiselev2003}. Easy-plane materials, such as Co and Co$_{90}$Fe$_{10}$ for the fixed layer and NiFe for the free layer, have  most often been pursued.  One drawback of utilizing materials with easy-plane anisotropy is that large (1 T) fields are often required to overcome the demagnetization fields needed to pull the magnetization out-of-plane.  

\subsubsection{All perpendicular}

Owing to their improved thermal stability, scalability, and potential for low to zero field operation, materials with perpendicular magnetic anisotropy (PMA) are extremely attractive.  One of the first \emph{all perpendicular} nanopillar STT devices combined a [Co/Pt]$_4$ multilayer as a fixed layer with a [Co/Ni]$_4$ multilayer free layer \cite{Mangin2006} and was primarily devoted to STT-induced switching.  Later STNOs with perpendicular free and fixed layers have shown zero-field operation \cite{Sim2012}.  

\subsubsection{Orthogonal}
	
In these types of devices, the equilibrium magnetization direction of the fixed and free layers are orthogonal to each other.  One of the first devices employed a perpendicular fixed layer based on a Co/Pt multilayer and a planar NiFe free layer \cite{Houssameddine2007}. More recent devices have tended to deal with the opposite situation, with in-plane fixed layers using either Co, Co$_{90}$Fe$_{10}$, or even NiFe, and a perpendicular Co/Ni multilayer free layer \cite{Rippard2010, Mohseni2011, Mohseni2013, Mohseni2014, Chung2014, Macia2014}.

\subsubsection{Tilted}
	
A tilted magnetization will have both in-plane and out-of-plane components; its use in STNOs is predicted to provide additional degrees of freedom to manipulate both static and dynamic states.  Relative to conventional in-plane or PMA magnetic films, the use of tilted fixed layers can optimize microwave signal generation and enhance spin-torque efficiency while maintaining high output powers with low to zero field operation \cite{Zhou2008, Zhou2009a, Zhou2009, He2010}.  To date, experimental work has primarily focused on the fabrication of tilted anisotropy exchange-coupled films by taking advantage of the competition between the PMA of a Co/Pd or Co/Ni multilayer and the in-plane shape anisotropy of a NiFe of CoFeB layer. \cite{Nguyen2011, Nguyen2012, Chung2013}.  Subsequent measurements have then focused on the fundamental spin wave excitations in such tilted anisotropy exchange springs \cite{Tacchi2013, Tacchi2014}.  Zero field operation in CoFeB/MgO/CoFeB MTJs with tilted CoFeB free layers has been experimentally demonstrated \cite{Skowronski2012}.

\subsubsection{Single layer}    
	
Departing from the trilayer configurations discussed above, theory \cite{Polianski2004, Stiles2004} predicts that spin filtering by a single FM layer can generate a sizable spin accumulation at each interface.  However, it is only when these interfaces are asymmetric that an imbalance of spin accumulation will lead to a significant STT.  Experiments on nanopillars with a single Co layer showed that sufficient STT could be generated to produce spin-wave excitations \cite{Ozyilmaz2004} and even switch the Co magnetization \cite{Ozyilmaz2006}.  The latest work on single layer devices has been in the nanocontact geometry, where evidence of a periodic vortex/antivortex creation/annihilation mode has been found \cite{Sani2013b}.

\subsection{STNOs: Magnetodynamical Modes}       

STNOs exhibit a diversity of fundamental spin wave modes that depend greatly on, for example, the device architecture, constituent FM materials, applied magnetic field strength, and applied magnetic field angle.  Here we summarize some of the most prevalent STT-induced excitations in both nanopillar and nanocontact STNOs.

\begin{figure}[tb]
  \begin{center}
    \includegraphics[ width=8.87cm]{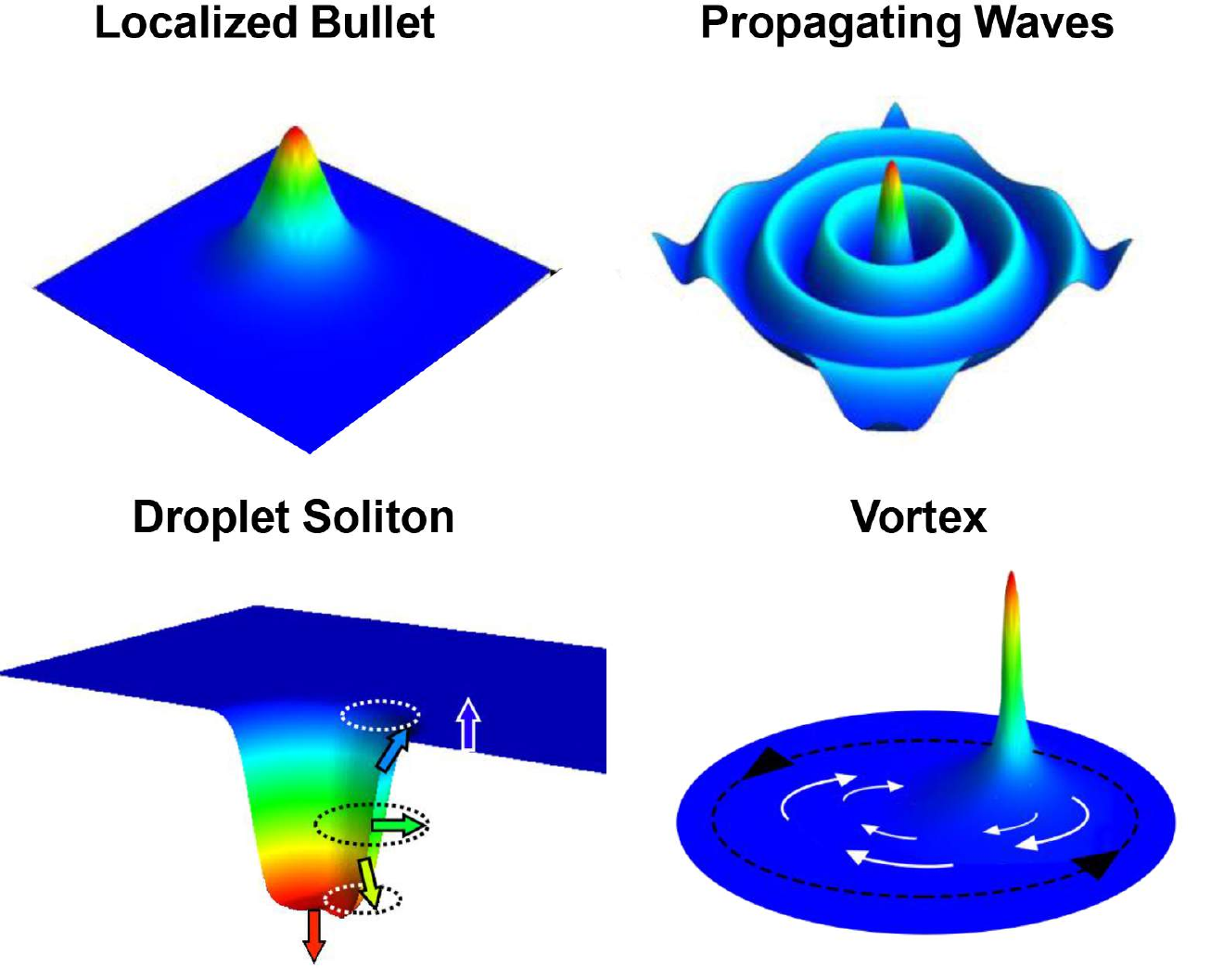}
    \caption{Fundamental spin wave modes.}
    \label{fig-modes}
  \end{center}
\end{figure}

\subsubsection{Nanopillar: Uniform Precession and Vortex Gyration}

The particular spin-wave mode generated in a nanopillar geometry depends most sensitively on the diameter or thickness of the free FM layer, as well as on the strength of the external magnetic field.  Frequently, the dimensions of the free layer are such that a magnetic vortex  characterized by an in-plane curling of the magnetization with a 10-nm sized out-of-plane vortex core at the center is a stable equilibrium configuration \cite{Cowburn2000}.  If so, then the STT will excite the fundamental gyration of the vortex core with frequencies typically less than 2 GHz \cite{Pribiag2007, Pulecio2014, Araujo2012, Araujo2013, Sluka2011, Locatelli2011, Khvalkovskiy2009, Bortolotti2013, Grimaldi2014}.  However, in a large enough external field, the vortex state will no longer be energetically favorable \cite{Lehndorff2009}, the magnetization will become more or less uniform, and STT will typically excite highly elliptical clam-shell precession in the case of a free layer magnetized in-plane.  Similarly, if either the diameter  or thickness of the free layer is sufficiently small that exchange interactions dictate a uniform magnetization, then such uniform precessions will also be generated.

\subsubsection{All in-plane nanocontacts: Localized Bullets, Propagating Spin Waves and Vortices} 

For nanocontact STNOs where the equilibrium magnetizations of both the free and fixed layers prefer to lie in-plane, the type of spin wave excitation depends strongly on the external applied magnetic field angle.  As described by both analytical \cite{Slavin2005} and micromagnetic simulations \cite{Consolo2007}, the initial experiments \cite{Rippard2004, Bonetti2010} using in-plane applied magnetic fields excited a strongly nonlinear, self-localized solitonic bullet mode.  However, when the free layer is magnetized perpendicularly to the film plane, as described by Slonczewski \cite{Slonczewski1999a}, an exchange dominated propagating spin wave with a wave vector inversely proportional to the nanocontact radius is excited, as directly proven using Brillouin light-scattering experiments \cite{Madami2011}.  For intermediate oblique angles, the behavior becomes much more complex; both simulations \cite{Bonetti2010} and experiments \cite{Bonetti2012} provide evidence of mode hoping \cite{Krivorotov2005a, Muduli2012, Heinonen2013} between propagating and localized bullet modes \cite{Dumas2013}. The use of elliptical nanocontacts has been useful in trying to disentangle the coupling mechanisms between these modes \cite{Iacocca2015}.

Special attention must be given in the nanocontact geometry to the critical role played by the current-induced Oersted field.  For typical drive currents and nanocontact dimensions, the Oersted field is on the order of 0.1 T at the nanocontact edge; this is a significant fraction of the external applied magnetic field and cannot therefore be considered an insignificant perturbation.  For example, the Oersted field results in a localization of the propagating mode for applied field angles near the film plane, which in turn promotes mode coexistence with the solitonic bullet mode \cite{Dumas2013}.  Furthermore, at large applied field angles, the Oersted field induces an asymmetric FMR field landscape which results in asymmetric spin wave propagation \cite{Madami2015}.  In short, in regions where the addition of the external field and the Oersted field produce a local field maximum, spin waves can no longer propagate, resulting in beam-like spin-wave propagation towards (and beyond) the local field minimum on the other side of the nanocontact \cite{Hoefer2008}. 

\begin{figure}[tb]
  \begin{center}
    \includegraphics[ width=8cm]{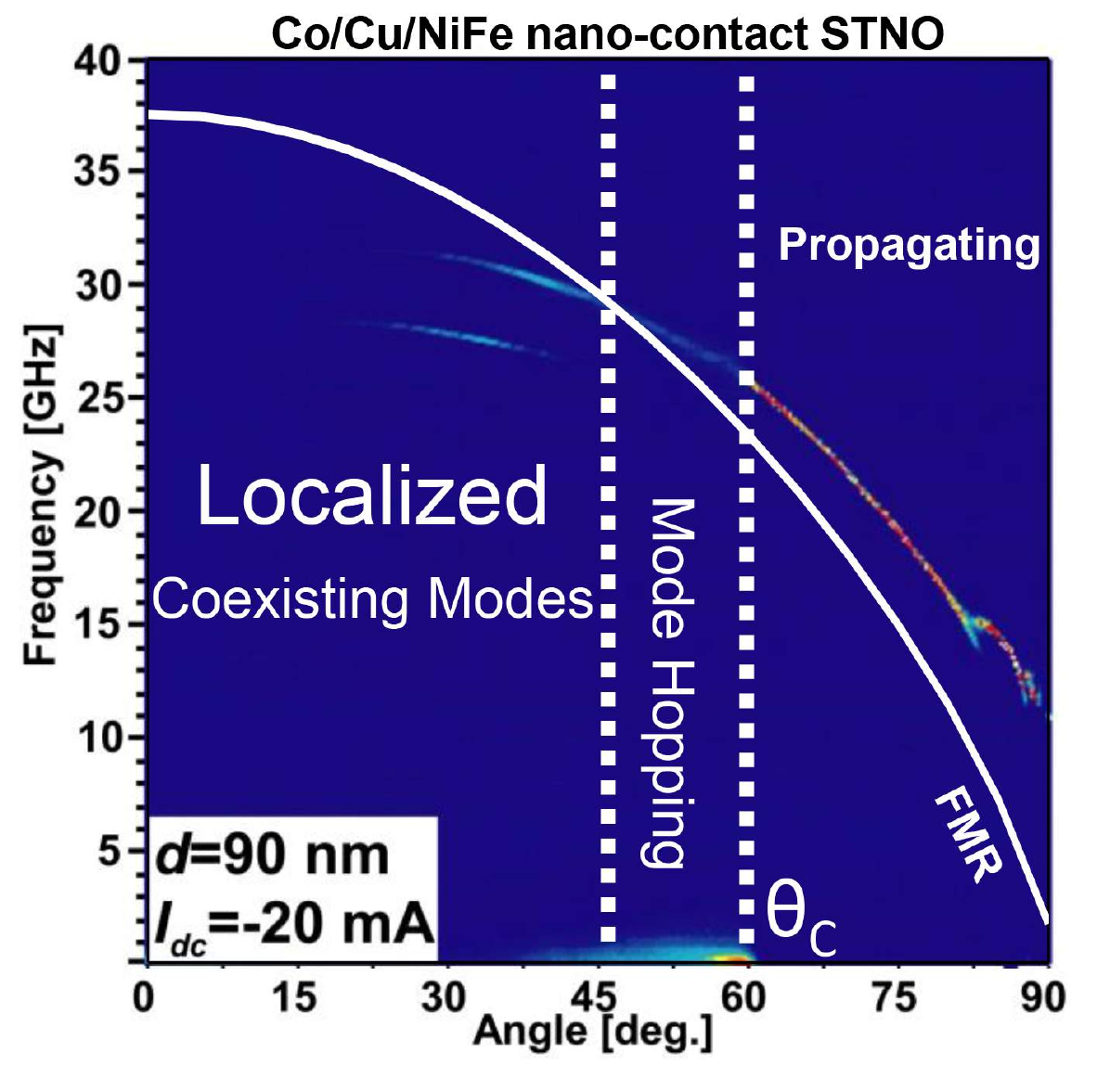}
    \caption{Angular dependence of spin wave modes in a Co/Cu/NiFe nanocontact STNO.}
    \label{fig-mod1}
  \end{center}
\end{figure}

Finally, the Oersted field also plays an important role in the formation of vortex oscillations in nanocontacts \cite{Pufall2007, Mistral2008, RuotoloA.2009, Manfrini2014}.  In small external magnetic fields, the circulating Oersted field favors a vortex state in the free layer via the Zeeman interaction.  This interaction in turn generates a confining potential for the vortex core.  The action of STT is then to drive the vortex core away from the center of the nanocontact, whereas the natural damping tends to pull the core back in.  In equilibrium, these two forces result in a self-sustained vortex gyration around the nanocontact.  In stark contrast to nanopillars, where the vortex motion is physically constrained, these vortex orbits can even occur in regions outside of the nanocontact itself.  In such cases, this results in a very large time-varying magnetoresistance signal and thus large output powers.  As with vortices in nanopillars, the oscillation frequencies are usually less than 2 GHz and the emission spectrum usually contains several higher harmonics due to elliptical orbits.  When driven past a critical velocity, the vortex core can flip polarity, leading to chaotic dynamics \cite{Petit-Watelot2012}.     

\subsubsection{Perpendicular and Orthogonal: Droplet Solitons}

Significant changes in the fundamental magnetodynamical modes are found when the free layer exhibits PMA, such as occurs in the oft-employed Co/Ni multilayers.  When the applied field is perpendicular to the film plane, three distinct operational modes can be identified.
At small applied magnetic fields, a small-angle precessional FMR-like mode is observed with relatively small power \cite{Rippard2010}.  As the magnitude of the drive current is increased (decreased), there is a gradual increase (decrease) in mode power (frequency) consistent with an increasing precession angle.  However, at larger applied fields, the oscillator behavior changes dramatically.  As the applied field is increased, a dramatic drop in oscillator frequency and an increase in oscillator power are observed, consistent with the formation of a magnetic droplet soliton \cite{Mohseni2013}. The theoretical basis for the droplet soliton was initiated by Ivanov and Kosevich \cite{Ivanov1977, Kosevich1990}, who showed  that the Landau--Lifshitz equation can sustain a family of conservative magnon drop solitons in the absence of damping.  While initially a theoretical curiosity, it was then shown that the balance between STT and damping underneath the nanocontact can provide the zero damping needed to sustain what are now called dissipative droplet solitons \cite{Hoefer2010, Hoefer2012}.  Finally, at very large external fields, the Zeeman energy will dominate and the droplet will collapse \cite{Dumas2014}.

\subsection{SHNO Device Architectures and Materials}

The influence of the spin current generated from SHE-induced spin accumulation on the ferromagnetic resonance of an adjacent magnetic layer has been initially shown on micrometer-sized permalloy discs \cite{Demidov2011a,Demidov2011b}. While the linewidth could be reduced by a factor of two, full compensation of the damping, and thus auto-oscillations, were not achieved in these micrometer-sized films, as the uniform application of the torque does not favor any auto-oscillation mode. SHNOs have thus been realized by nanopatterning of various parts of the bilayer to achieve the nonuniform STT, and can be categorized into four primary device geometries. In the three-terminal nanopillar geometry, the magnetic free layer is a nanosized element on top of an extended slab of the SHE-generating normal metal \cite{Liu2012,Liu2012prl}. The electrical read-out is carried out here through an MTJ element including the oscillating free layer, which is how appreciable output powers can be achieved. Another approach is to utilize a nanogap between two highly conductive electrodes, injecting the necessary high current densities into a nanosized region of an extended FM/Pt bilayer \cite{Demidov2012}. In the more recent nanoconstriction-based SHNOs, both the FM and the Pt layer are patterned into a nanoscopic constriction \cite{Demidov2014}. Similar confinement of the SHE-induced spin current has been demonstrated in a 190 $\mu m$ nanowire geometry \cite{Duan2014}.  The benefits of the latter two SHNO device geometries over their STNO counterparts include easier fabrication (for example, they require the deposition of only two layers and fewer lithographic steps) and %provide 
direct optical access to the oscillating region. In contrast to STNOs, which rely on the GMR or TMR effects, these SHNOs must however rely on the smaller anisotropic magnetoresistance (AMR) effect in a single FM layer, and therefore typically exhibit lower output powers.  

%SHNOs come in two primary device geometries, utilizing either a nano-gap \cite{Demidov2012} or nano-constriction \cite{Demidov2014}, to focus the charge current, and therefore the resulting SHE generated pure spin current, to a nanoscopic region.  In the nano-gap devices pointed gold electrodes are placed on top of an extended FM/Pt bilayer whereas in nano-constriction devices both the FM and Pt layers are patterned into a nanoscopic constriction.  Benefits of SHNOs over their STNO counterparts include easier fabrication, e.g. they require the deposition of only two layers and fewer lithographic steps, and provide direct optical access to the oscillating region.  In contrast to SHNOs which rely on the GMR or TMR effects, SHNOs must rely on the much smaller anisotropic magnetoresistance (AMR) effect in a single FM layer and therefore typically exhibit much lower output powers.  

The generation of a sufficient SHE-induced pure spin-current requires a material with a large spin-orbit coupling. To date, only results on SHNOs using Pt and Ta have been published. However, a few different FM materials have been studied in addition to NiFe, including Co$_{40}$Fe$_{40}$B$_{20}$ \cite{Ranjbar2014}, Co/Ni multilayers with PMA \cite{Liu2015}, and the ferrimagnetic insulator yttrium iron garnet (YIG) \cite{Collet2015}.  

\subsection{SHNOs: Magnetodynamical Modes}

%The majority of published experimental results on SHNOs have been with in-plane applied fields.

For in-plane magnetized thin films, two different auto-oscillation modes have been observed in nanogap SHNOs. As described in \cite{Demidov2012}, these devices have been studied for in-plane applied fields with microfocused Brillouin light scattering and from  \emph{(i)} the sub-FMR frequency and \emph{(ii)} the spatial size, which was smaller than the instrument resolution, the existence of a self-localized spin wave bullet was inferred \cite{Slavin2005}. Electrical measurements as a function of the temperature, however, showed the existence of the low-frequency and low-linewidth bullet mode only for cryogenic temperatures with a second high-frequency mode appearing at higher temperatures, and even being dominant at room temperature \cite{Liu2013}. Micromagnetic simulations suggest that this high-frequency mode is  generated within local field minima caused by the bullet mode---which implies the simultaneous excitation of both \cite{Ulrichs2014}. However, since the Oersted field generated from the nonuniform current in the Pt layer always counteracts the applied in-plane field, a similar field minimum in the nanogap is always present, allowing the high-frequency mode to form. Using a 1~$\mu$m large gap between the electrodes, and thus generating rather uniform current distributions, \cite{Collet2015} showed that the high-frequency mode most likely resembles a coherent excitation of the FMR mode.  In the nanowire SHNOs, the existence of both bulk and edge modes has been identified \cite{Duan2014}

While the linear high-frequency excitation shows an increasing threshold current with the strength of the applied field, the spin wave bullet has a relatively constant threshold current as long as the internal magnetization angle stays below a critical value \cite{Gerhart2007}. The output characteristics can be thus tuned between these two regimes via the application of a tilted magnetic field, where the electrically measured linewidth is lowest for high fields, since only the spin-wave bullet exists there \cite{Durrenfeld2015a}. Moreover, the interplay can be varied by the nanogap size, favoring the bullet mode for smaller sizes. When both modes are energized at the same time for intermediate field strengths, their linewidths are broadened due to mode coupling events \cite{Iacocca2015}.
%, which have been further studied by electrical measurements, micro-focused Brillouin light scattering, or micromagnetic simulations. The signature of

Propagating spin waves with frequencies above FMR are important in, for example, mutual synchronization, and are predicted by micromagnetic simulation in nanogap SHNOs for out-of-plane fields  \cite{Giordano2014a}. However, in this study, the current densities required are about twice as large as for the in-plane field modes, which inhibits the experimental observation of the propagating spin wave modes.

Nanoconstriction-based SHNOs show a mode that evolves from the thermal FMR, as shown by micro-focused BLS measurements and electrical detection \cite{Demidov2014}, and is therefore comparable to the high-frequency mode of nanogap SHNOs. However, with the constricted geometry around the oscillation center, the presence of %a second mode, 
the bullet is hindered and the single-mode excitation can be detected with lower linewidths. %Similarly to the nano-gap SHNOs, the application of a tilted magnetic field can alter the oscillation behavior towards a self-localized bullet mode, exhibiting even lower linewidths \cite{Awad2015}.

%To date all published results on SHNOs have been with in-plane applied fields. For NiFe and Co$_{40}$Fe$_{40}$B$_{20}$ free layers localized spin wave bullet modes have been observed either directly using micro-focused Brillouin light scattering \cite{Demidov2012} or by purely electrical means \cite{Ranjbar2014, Liu2013}.
%Besides the non-linear spin wave bullet mode for in-plane magnetized films, also the generation of a coherent FMR-like precession has been shown for nano-gap SHNOs \cite{Liu2013,Collet2015}.  
%Finally, evidence of a unique dynamic skyrmion have been found in SHNOs with Co/Ni free layers \cite{Liu2015}.

%\textbf{Ahmad/Philipp: Should we add here text related to our unpublished results using OOP fields?}

\subsection{Mutual Synchronization}

Synchronization to external sources, as described later in regards to injection locking, or mutual phase locking of individual STOs, as described here, are techniques often used to improve  oscillator power and linewidth. There are several coupling mechanisms that have been investigated to synchronize STNOs to each other, each with a varying degree of experimental success.  Synchronization of serially connected nanopillar STNOs using their shared microwave electric current \cite{Grollier2006} has been theoretically proposed, but to date has shown no clear success.  Another route to synchronize nanopillars is to utilize the relatively long-range dipolar stray microwave fields generated during precession. Recent simulations show this to be a feasible solution for vortex-based oscillators \cite{Araujo2015}.  

To date, however, the most successful experimental works on synchronizing STNOs have occurred in the nanocontact geometry.  In such devices, the nanocontacts are patterned on top of a shared free layer, which in turn provides the coupling medium.  The first works to show that two STNOs could be synchronized were published in 2005 \cite{Kaka2005, Mancoff2005}.  It was soon shown experimentally \cite{Pufall2006}, as well as investigated theoretically \cite{Slavin2006, Victora2009} and micromagnetically \cite{Kendziorczyk2014}, that the coupling mechanism is mediated by propagating spin waves.  Since these early experiments, progress has remained relatively slow.  Only in 2009 was it demonstrated that four vortex-based STNOs could synchronize, this time by direct exchange interactions mediated by an antivortex. \cite{RuotoloA.2009}. Then, in 2013, the successful synchronization of three high-frequency STNOs was shown for the first time \cite{Sani2013a}.  Recent simulations have shown robust synchronization of an arbitrary number of nanocontact vortex oscillators mediated by the strong magnetodipolar interaction \cite{Erokhin2014}. 

The most recent work on synchronization has focused on the use of highly directional spin-wave beams. While it has been shown that the Oersted field generated in the vicinity of the NC can dramatically alter the emission pattern of SWs \cite{Hoefer2008, Madami2015, Dumas2013}, its role in the synchronization behavior of multiple NCs has been largely ignored.  Synchronization is promoted (impeded) by the Oersted field landscape when the nanocontacts are oriented vertically (horizontally) with respect to the in-plane component of the external field, due to the highly anisotropic spin-wave propagation.  Not only is robust synchronization between two oscillators observed for separations greater than 1000 nm, but synchronization of up to five oscillators has been observed in the vertical array geometry.  Furthermore, the synchronization must be considered driven, rather than mutual, in nature as the final frequency is enforced by the STNO from which the SW beam originates \cite{Houshang2015}. 

%\textbf{Ahmad/Philipp: Should we add here text related to our unpublished results synchronizing SHNOs?}

\section{Functional properties}

%Probably the place to describe synchronization, maybe of both STNOs and SHNOs. Possibly the introduction of the spin torque diode to match Tingsu's discussion about applications.

\subsection{Frequency range}

%\subsubsection{Nano-contact STNOs}

The precession frequency of the STNOs is tunable over a large range, both through the direct current and the applied magnetic field. The  precession  frequency generally increases as the external magnetic field  increases. Its behavior is generally governed by the Kittel equation, according to which the precession frequency increases as the square root of the applied field at 
low fields and linearly at high fields. Theoretical works predict a very high frequency of 0.2 THz for the common ferromagnetic material Ni$_{80}$Fe$_{20}$.~\cite{Hoefer2005} In the very first work by  Kiselev \textit{et al.},~\cite{Kiselev2003} a highest frequency of ~25 GHz was demonstrated. The devices used were nanopillars of Co/Cu/Co/ multilayer. A year later, Rippard \textit{et al.}~\cite{Rippard2004} demonstrated a frequency of  38 GHz in a nanocontact device based on a Ni$_{80}$Fe$_{20}$ free layer. The field tunability of frequency is 18.5 MHz/mT and the current tunability is about -200 to 400 MHz/mA for  nanocontact STNOs.~\cite{Bonetti2009, Bonetti2010, muduli2011if}  Bonetti \textit{et al.}, ~\cite{Bonetti2009} reported  a frequency as high as 46 GHz for high magnetic fields applied normally to the film plane. The STNO frequency was limited by instrumentation. It was reported that it would be possible to achieve a frequency beyond 65 GHz, which would be  potentially attractive for a range of millimeter-wave applications, such as short-range high-speed radio links (IEEE 802.15.3c) near 60 GHz and vehicle radar for active cruise control at 77 GHz.~\cite{Hasch2012}

%\subsubsection{Nanopillar STNOs}

\subsection{Microwave output power}

 An STNO may be viewed as a time-dependent resistor with its resistance given by:
 
\begin{equation}
R=R_{\rm av}+(\Delta R/2)\cos(\omega t)
\label{eq:pow}
\end{equation}
where $R_{\rm av}$ is the average resistance of the STNO, $\Delta R$ is the magnetoresistance of the material stack, and $\omega$ is the precession frequency. For a GMR-based device, $\Delta R$ is low, and hence the power generated by the GMR-based STNOs is also low, typically in the sub-nW range.~\cite{Kiselev2003,Rippard2004}  Higher powers have been achieved in magnetic tunnel junction based material stacks, for which $\Delta R$ is significantly larger. In magnetic tunnel junctions, powers of up to 3.6 $\mu$W have already been reported for spin-torque vortex oscillators with an FeB free layer \cite{Tsunegi2014}. However, the frequency of operation of these vortex oscillators is typically low. For uniform mode oscillation, the highest power achieved so far has been 2 $\mu$W, obtained from a  sombrero-shaped nanocontact magnetic tunnel junction based STNO with a perpendicular anisotropy free layer \cite{Maehara2013}. For in-plane magnetized MTJ-STNOs, the highest power reported was 140~nW \cite{Deac2008} for a frequency around 5 GHz. While the frequency can be increased using the magnetic field, the output power usually drops with increasing field strength;  obtaining higher power at higher frequencies is thus a challenge. A novel method of achieving this involves frequency doubling in MTJ-STNOs \cite{muduli2011jap}. In that work, powers of up to 1.25 nW were demonstrated at a frequency of $\sim$10 GHz.

\subsection{Modulation properties}

\subsubsection{Modulation of uniform mode STNOs}

\begin{figure}[tb]
  \begin{center}
    \includegraphics[ width=8.87cm]{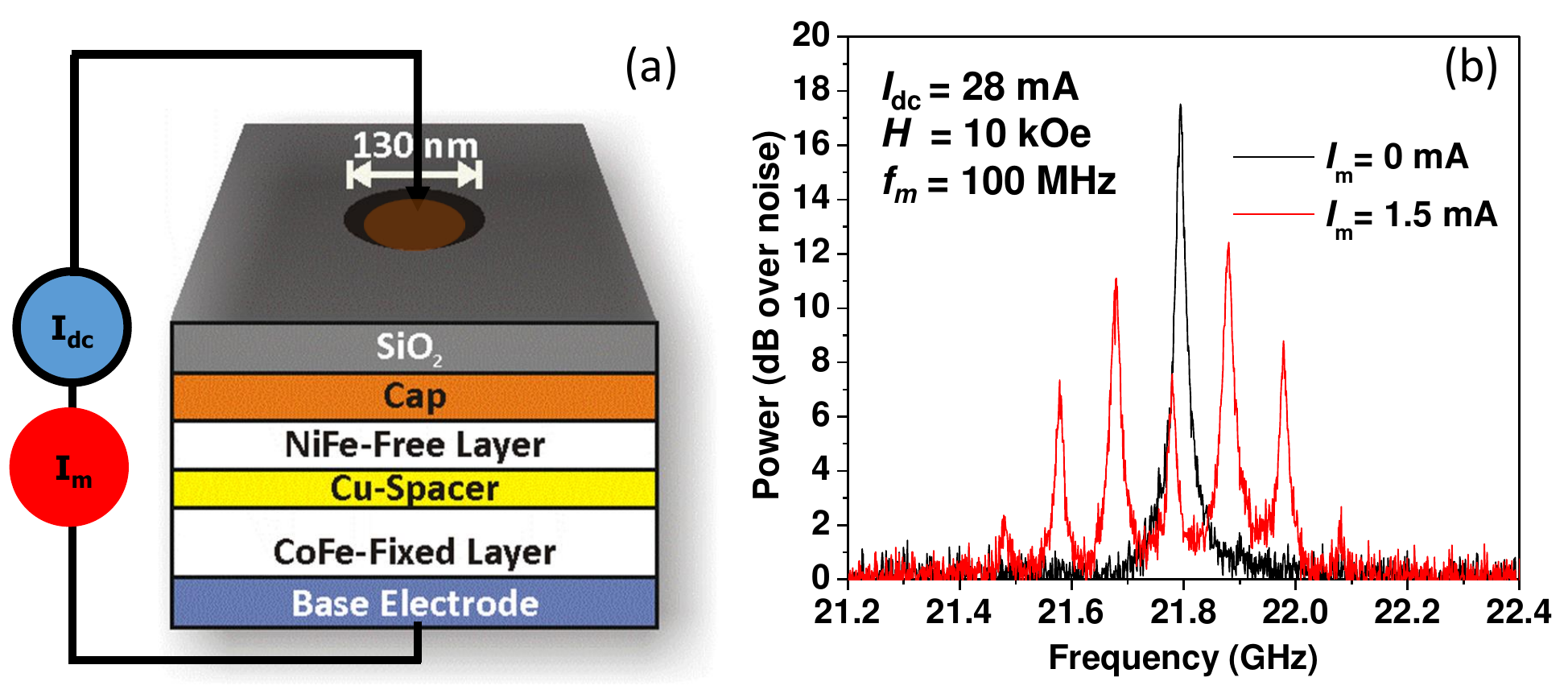}
    \caption{(a) A nanocontact STNO biased with dc and modulating RF currents ($I_{\rm m}$). (b) Sample spectra of the device with and without modulation.}
    \label{fig-mod1}
  \end{center}
\end{figure}
Modulation is a very important requirement for the communication applications of STNOs. Modulation studies of STNOs are, however, surprisingly limited despite the issue's critical importance for  communications applications. The first experimental demonstration of frequency modulation in an STNO was reported in 2005~\cite{pufall2005}, with the demonstration that the addition of an alternating current to the direct current results in  frequency-modulated spectral output, generating sidebands spaced at the modulation frequency. This phenomena is demonstrated in Fig.\ref{fig-mod1} (b), where equally spaced multiple sidebands are produced upon the addition of an RF current (1 mA, 100 MHz). The device is shown in Fig.\ref{fig-mod1} (a) and corresponds to that of \cite{muduli2010}. As can be seen in this figure, the sidebands have unequal amplitudes and the carrier frequency shifts with the addition of the RF current. This phenomena was first reported in ~\cite{pufall2005} and was attributed to nonlinear frequency modulation. However, in 2010, it was shown that the unequal sideband amplitudes result from a combination of nonlinear frequency and amplitude modulation (NFAM).  It was shown that the combined nonlinear frequency and amplitude modulation model can accurately describe the experimental data without any adjustable parameters. The frequency of modulation was 40 MHz in \cite{pufall2005} and 100 MHz in \cite{muduli2010}. Later,  it was shown that a nanocontact-based STNO can be successfully modulated to a modulation frequency of 3.2 GHz \cite{muduli2011if}. The agreement of experiment with the NFAM model is maintained until 1 GHz \cite{muduli2011ieeem}. It was also shown that modulation can be used to narrow  the linewidth of STNOs by as much as 85 \%~\cite{pogoryelov2011a}. This linewidth reduction is due to a general averaging effect of modulation on the nonlinearity of the STNO signal and was explained on the basis of nonlinear auto-oscillator theory \cite{Slavin2009a} and nonlinear frequency modulation.

\begin{figure}[tb]
  \begin{center}
    \includegraphics[ width=8.87cm]{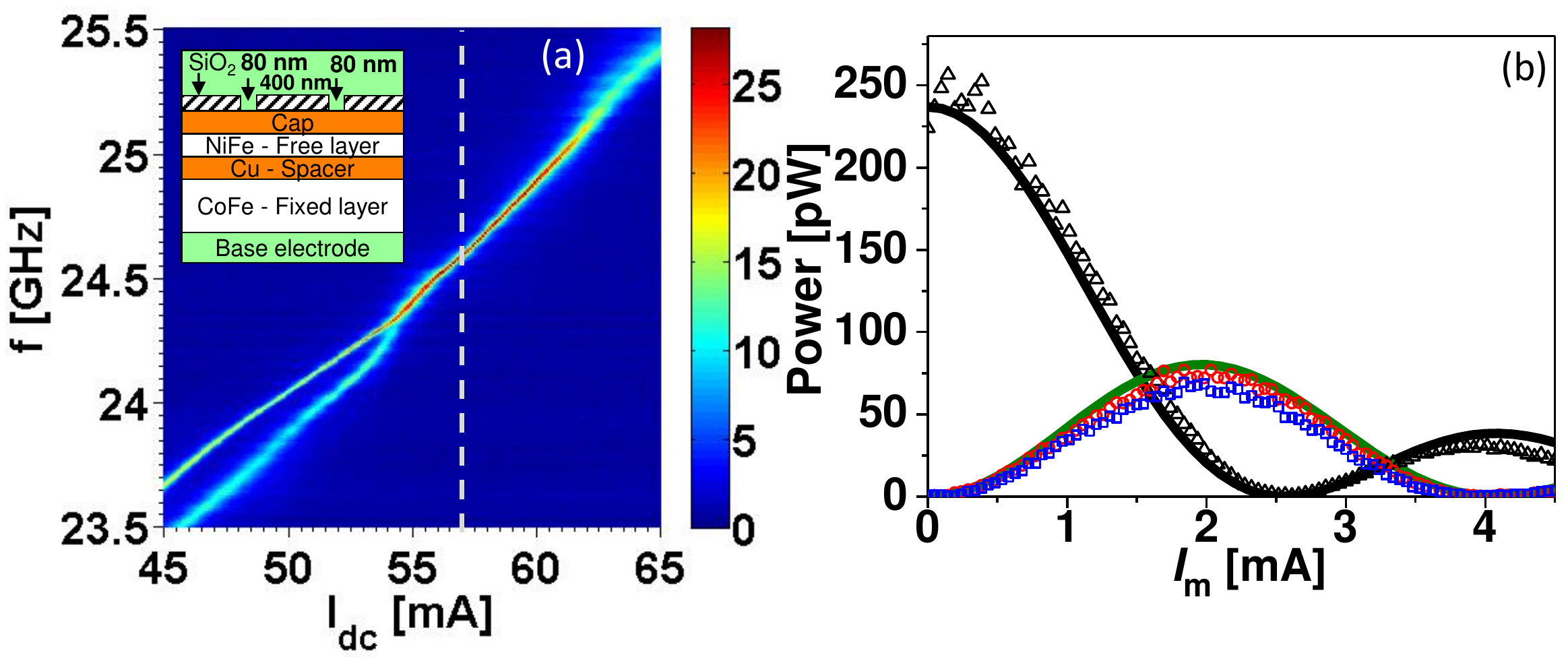}
    \caption{ Modulation of a pair of nanocontact STNOs from \cite{pogoryelov2011b}.  (a) Map of  peak power versus frequency and applied current bias. The inset shows a schematic cross-section of the two nanocontacts on a GMR spin-valve mesa. (b) Experimental and calculated values of integrated power for $I_{dc}=57$ mA. The solid lines represent the corresponding calculated integrated power from NFAM theory.}
    \label{fig-mod2}
  \end{center}
\end{figure}

In 2011,  modulation of a pair of nanocontact STNOs was successfully demonstrated\cite{pogoryelov2011b,muduli2011aip,muduli2011ieeem}.  In Fig.\ref{fig-mod2}, we reproduce the results from \cite{pogoryelov2011b} which show that the synchronized state of a pair of nanocontact STNOs (at 57 mA) can be maintained while modulating. Furthermore, the modulation behavior can be explained by assuming a single oscillator model, as shown in Fig.\ref{fig-mod2}(b). This finding was very promising for applications involving a large array of synchronized STNOs. Many electronic circuits use  injection-locked phase-locked loops (PLLs) as one of the building blocks of RF transceivers and so it is also crucial for applications to demonstrate the modulation of the injection-locked state of the STNO. More recently, modulation of injection-locked MTJ-STNOs was demonstrated and the  modulation-mediated nonresonant unlocking mechanism was described \cite{Durrenfeld2014b}. Interestingly, the unlocking phenomena sets a lower limit on the modulation frequency but no upper limit on the modulation frequency for injection-locked MTJ-STNOs.

\subsubsection{Modulation of vortex STNOs}
Experimental studies of the modulation of vortex-based STNOs are rather limited. The first digital modulation of vortex-based STNOs was demonstrated in \cite{manfrini2009apl}. Using frequency shift keying (FSK), it was shown that the oscillator can switch between two stabilized frequencies differing by $25\%$ in ~20 ~ns. This performance is difficult to achieve in a voltage-controlled oscillator, and  vortex-based STNOs hence hold much promise for FSK modulation-based applications. More recently, analog frequency modulation has been demonstrated in a magnetic tunnel junction nanopillar in in-plane fields~\cite{Martin2013}. In this work, the authors show modulation for modulation frequencies that are up to 10\% of the oscillator frequency. In these experiments, however, the modulation period is comparable to the characteristic time of the vortex dynamics, and hence the frequency modulation data was modelled using ``deviation sensitivity,'' which differs significantly from the tunability measured in a quasistatic regime.

\subsubsection{Modulation rate}

When the modulation frequency of an STNO is increased, the STNO may not be perfectly modulated. A recent study has shown that the upper limit of the frequency of modulation in an MTJ-STNO is ~200 MHz~\cite{Quinsat2014}. In this work, the authors showed that the bandwidth of modulation is limited by the amplitude relaxation rate $\Gamma_{p}$ and that the power level of  amplitude modulation strongly decreases beyond $f_{m}>f_{p}$, where  $f_{p}=\Gamma_{p}/\pi$. This finding is  in agreement with the nonlinear auto-oscillator theory of STNOs \cite{Slavin2009a}. Although the original nonlinear auto-oscillator theory does not include amplitude modulation, a recent theoretical study that includes  an NFAM scheme also suggests that the modulation bandwidth of STNOs is determined by $\Gamma_{p}$\cite{Iacocca2012}.

%%%%%%%%%%%%%%%%%%%%%%%%%%%%%%%%%%%%%%%%%%%%%%%%%%%%%%%%%%%%%%%%%%%%%%%%%%%%%%%%
\subsection{Phase noise}

The miniature size of STNOs and SHNOs compared to industry-standard oscillators inevitably makes their oscillation more susceptible to noise. The stability of any oscillator can be characterized in terms of its amplitude noise and phase noise, where the width of the spectral peak (commonly referred to as the linewidth) is usually governed by the phase noise---as has also been found for STNOs. The phase noise of SHNOs has not been as extensively studied as that of STNOs, but one initial report \cite{Liu2013} provides a lowest linewidth on par with what has been achieved for STNOs.

\subsubsection{Oscillation modes and mode-hopping}
A prerequisite for stable oscillation is that the oscillation is not intermittent. One type of intermittency is on--off toggling, which has been observed for low drive currents \cite{Houssameddine2009}. Another type is mode-hopping between different distinct frequencies. STNOs do in many cases show spectra with the presence of several modes represented by distinct frequencies, where time-domain measurements can often reveal a random mode-hopping between these frequencies. The mode-hopping rate can range from milliseconds down to nanoseconds, meaning that, depending on the bandwidth of the characterization setup, it might not always be observed. It should be noted that in some cases there is the possibility of actual coexistence (in time), a  clear indicator of which is the appearance of second-order intermodulation products \cite{Dumas2013}. If the application is low bandwidth compared to the mode-hopping rate, the mode-hopping itself need not constitute too large  a problem, since the fast fluctuations might be washed out.

However, the mode-hopping has a large negative effect on the coherence of the oscillation. In a mode-hopping event from one frequency to another and then back again, the phase will be lost (introducing phase noise). This becomes visible in the spectrum in the form of a dramatic linewidth increase \cite{Bonetti2012prb,Iacocca2014b}.

The occurrence of multiple simultaneously excitable modes is  most evident in nanopillar STNOs \cite{Muduli2012a}, where the modes emerge  as various edge and center modes on account of its confined geometry. In nanocontact STNOs, with their far-extending magnetic films, mode-hopping within the propagating mode has been experimentally observed only at very high drive currents (close to the breakdown of the oscillation),~\cite{Eklund2013} while simultaneous excitability of the propagating and localized bullet mode can also induce mode-hopping \cite{Bonetti2010, Dumas2013}. Moreover, the propagating mode in nanocontact STNOs generally exhibits discontinuous mode transitions as a function of the bias point \cite{Rippard2006,Pufall2012,Muduli2012b}. Close to these transitions, the linewidths of the individual modes typically increase \cite{Sankey2005,Muduli2012b}, which could be theoretically explained by rapid  hopping between the modes \cite{Iacocca2014b}, though this has not been experimentally confirmed yet.

\begin{figure}[tb]
  \begin{center}
    \includegraphics[ width=8.87cm, trim=0cm 0cm 0cm 0cm, clip=true,]{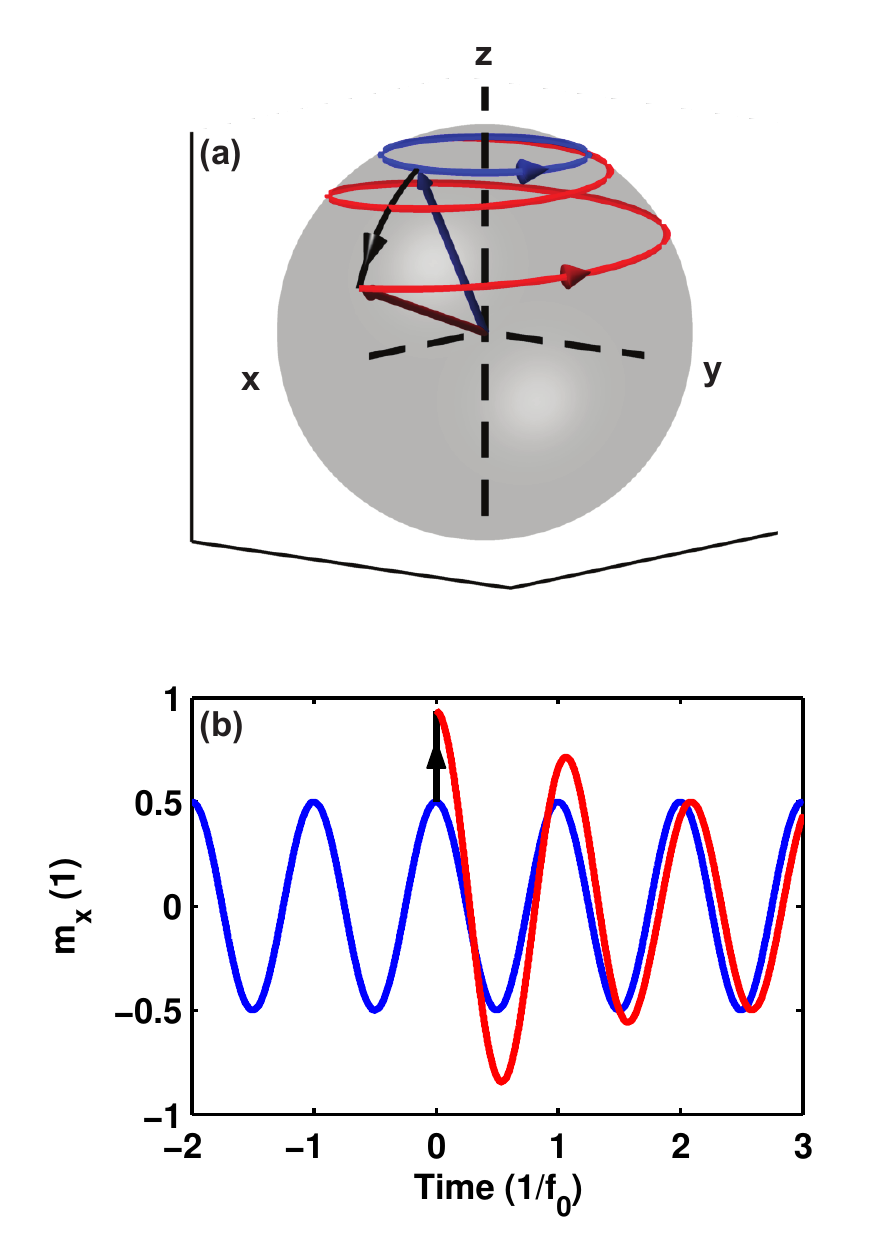}
    \caption{Phase noise caused by an amplitude perturbation (black) at $t = 0$ for an oscillator at the steady-state frequency $f_0$ with decreased frequency at higher oscillation amplitude. (a) Trajectories of the unperturbed (blue) and perturbed (red) precession. (b) Magnetization projection onto the $x$ axis, showing the resulting phase shift.}
    \label{fig-phasenoise}
  \end{center}
\end{figure}

\subsubsection{Single-mode phase noise}
Apart from control over the existence of multiple oscillation modes (and the instability they introduce to each other), a second fundamental question is that of the stability of a single mode. This stability is primarily considered to be governed by thermal magnetic noise which is usually modelled as random fluctuations in the magnetic field. This has a straightforward implementation in the macrospin approximation. Micromagnetically, it is usually implemented as fluctuations that are uncorrelated in both time and space \cite{BerkovHMM2007}. It has been noted \cite{Silva2010}, that no rigorous theory has been presented to explain the effects of spatial fluctuations on the spin-wave eigenmodes.

Another aspect of stability arises from the complexity of micromagnetic systems, where spatial inhomogeneity can be induced by locally varying fields even at zero temperature \cite{Lee2004,BerkovPRB2005a,BerkovJMMM2007}. Such spatially varying fields may stem from the geometry through the fabricated shape of the free layer or the roughness in the nanometer-scale film thickness, or through local anisotropy fields in the metal grains.

The noise properties have been found to be highly dependent on the strong amplitude--frequency coupling \cite{Bianchini2010} which acts to transfer amplitude noise into frequency noise \cite{Kim2008a,Silva2010}. As the magnetization is precessing around its steady-state orbit, fluctuations act to displace it along two orthogonal directions: in parallel with and perpendicularly to the orbit. Parallel displacement introduces a single phase shift whereas a perpendicular perturbation (Fig. \ref{fig-phasenoise}(a)) provides no initial phase difference. However, since a perpendicular perturbation places the magnetization away from the steady-state orbit at an oscillation amplitude different from the steady-state amplitude, the magnetization will momentarily take on another frequency associated with this new amplitude. During the time of precessional relaxation of the amplitude back to the steady-state orbit, the frequency will gradually approach the steady-state frequency, but before this frequency has been reached, the oscillator will have accumulated a phase shift as compared to the nominal frequency (Fig. \ref{fig-phasenoise}(b)).

Quantitatively, both theoretical calculations \cite{Kim2008a,Silva2010} and experiments \cite{Georges2009,Quinsat2010a,Sierra2012} show that the main effect of perpendicular perturbations and of amplitude-frequency coupling is by far the dominating mechanism for the phase noise in STNOs. In the framework of the general nonlinear auto-oscillator theory, it has also been demonstrated how, in the vicinity of the oscillation threshold,  the linewidth increases due to spectral line asymmetry induced by several statistical solutions to the noise process \cite{Kim2008b}.

The high impact of the strong amplitude--frequency coupling on the resulting phase noise means that the oscillator will be least sensitive to fluctuations when its amplitude--frequency coupling is nonexistent, but can also be improved by achieving a higher amplitude relaxation rate \cite{Lee2013}. From a fundamental thermal noise perspective, the frequency stability should be increased by the use of materials with low thermal damping of the magnetization precession---i.e. materials that exhibit a low value for the Gilbert damping parameter. On the contrary, however, the amplitude relaxation rate is proportional to the Gilbert damping, meaning that the amplitude perturbations will be damped out more slowly. Since the amplitude fluctuations are the major source of frequency noise, a high value for the Gilbert damping parameter is preferable. \cite{Silva2010}.

\subsubsection{1/f noise}
Several experimental studies report the presence of low-frequency $1/f$ frequency noise in both MTJ nanopillar \cite{Quinsat2010a,Sharma2014} and GMR nanocontact STNOs \cite{Keller2010,Eklund2014}. From a characterization point of view, the high level of $1/f$ frequency noise is challenging in the sense that the standard phase noise measurement modes of spectrum analyzers are unable to lock onto the STNO electric RF signal. The research community has been able to overcome the characterization problem through time-domain measurement of the waveform and subsequent digital signal processing \cite{Keller2009}, but from an application point of view, $1/f$ noise remains an issue. This $1/f$ frequency noise has been reported to increase in the presence of multiple oscillation modes \cite{Eklund2014,Sharma2014}, but is nonetheless of considerable strength in the apparently single-mode case. There is currently no theory of STNOs that can  account for the presence of  low-frequency $1/f$ frequency noise in single-mode oscillations, but it has been suggested that this might be related to the existence of unexcited oscillation modes or $1/f$-type amplitude noise possibly introduced by microscopic material inhomogeneity \cite{Eklund2014}.

%\subsubsection{Synchronization}
%For increased coherence, mutual %synchronization of two or more STNOs has been %proposed. The two initial experimental reports %of mutual synchronization of two nano-contact %GMR STNOs do indeed show a decreased linewidth %as the two oscillators phase lock to each %other. I may want to raise the question of how %much the linewidth improves in relation to the %increased size of the total oscillator system; %twice the oscillating volume should in itself %give about a factor of two linewidth %improvement.

%\textbf{Randy Text:}

%A significant, and generally accepted, drawback of STNOs is their relatively low output power, typically in the pW range for all-metallic nanocontact STNOs and in the $\mu$W range for the best MTJ-based nanopillar devices.  In addition to oscillator power, the high phase noise, particularly problematic for MTJ nanopillar devices, is an additional hurdle that must be overcome. The high phase noise of STNOs is primarily due to the relatively small excited mode volumes, which are then susceptible to device imperfections and thermal noise.  One of the most pursued methods to both increase the oscillator power and to decrease the phase noise has been to synchronize many oscillators together.  Theoretically, for perfect in-phase mutual synchronization, the output power should scale as $N^2$ (for serially connected oscillators) and the linewidth should at the same time decrease substantially. 

\subsubsection{Injection locking and phase-locked loops}
Two major paths are currently discernible  for stabilizing the generation of STNO and SHNO frequencies by applying  electrical signals  to the device. The first  is  injection-locking, in which the STNO \cite{Rippard2005,Sankey2006,Georges2008,Lehndorff2010,Urazhdin2010,Urazhdin2010b,Tabor2010,Dussaux2011,Rippard2013,Bortolotti2013,Durrenfeld2014,Lebrun2015} or SHNO \cite{Demidov2014b} is locked to an injected RF signal at the same nominal frequency or at an integer or integer fraction multiple thereof. The locking can be achieved through direct electrical injection of the external RF signal (superposed on the DC drive current) or through an AC magnetic field from an adjacent microstrip line. One compelling system is that in which the STNO/STHO itself acts as the injection source, thereby acting to stabilize itself. Such a system has been investigated \cite{Khalsa2015} analytically and, for the case of a vortex MTJ STNO, also numerically. With the correct choice of delay time for the reinjected signal, the linewidth for this vortex MTJ STNO case could be decreased by a factor of 4. Self-injection locking has recently been achieved also experimentally, showing a maximum decrease in linewidth by a factor of 2.5 \cite{Tsunegi2015}.

A second path is to use a phase-locked loop (PLL) for the STNO/SHNO. In a PLL, the oscillator's RF signal is compared with a reference signal and the resulting phase difference is used to instantly rebias the oscillator, taking advantage of its frequency-tuning mechanism. PLL-locking through an external reference signal has been experimentally achieved for STNO vortex oscillations below 150 MHz \cite{Keller2009} and, very recently, for a MTJ STNO at 5 GHz \cite{TamaruINTERMAG2015}. The limited number of reported studies is probably  related to the lack of PLL circuits with bandwidths high enough to track the STNO signal.

\section{STNO- and SHNO-based applications}
A unique feature of STNOs and SHNOs is their ability to provide multiple functionalities, from microwave source to microwave detection and signal processing. This versatility opens up the possibility of employing STNOs and SHNOs in a wide variety of applications. The most common applications include microwave sources, microwave detectors, noncoherent transceiver, magnonics, neuromorphic computing, magnetic field sensing, and magnetic field generation. These applications will be introduced in this section. Moreover, the issues, challenges, and developments in STNO- and SHNO-based applications will be discussed.

\subsection{STNO-based applications: the state of the art and prospects} 

\subsubsection{Microwave source}
The miniature size, very wide tunability, high integration level with CMOS technology, and low power consumption of STNOs make them good candidates for  current-controlled oscillators in multiband multistandard radio systems \cite{Villard2010a, Chen2014a}. Table I summarizes the performance of different STNOs in the literature. As can be seen from Table I, a single MTJ STNO \cite{Maehara2014, Muduli2011a} can be used as a local oscillator (LO) covering the entire bandwidth of ultrawideband (UWB) applications. GMR STNOs typically have higher and broader tunable frequency ranges. The operating frequency of GMR STNOs has been experimentally observed up to 46 GHz and extrapolated to 65 GHz \cite{Bonetti2009}. It has been theoretically estimated that the operating frequency of STNOs may approach 200 GHz if STNO nanocontacts can be made much smaller \cite{Hoefer2005}. More recently, a very low Gilbert damping and ultrafast spin precession with frequencies of up to 280 GHz has been demonstrated \cite{Mizukami2011}, which makes it possible to further extend the operating frequencies of STNOs. The extremely high and broad operating frequency range of STNOs also opens up the possibility of utilizing them in millimeter wave application---for instance, in wideband data transmission systems. 

\begin{center}  
\begin{table*}[tb]
 \caption{STNO summary}
\label{table:kysymys}
\centering
\small
\begin{threeparttable}
\begin{tabular}{| p{2.45cm} | p{1.45cm} | p{1.76cm} | p{1.4cm}| p{1.9cm} | p{1.25cm}| p{1.85cm}| p{2.2cm}| }
\hline\hline
Device& Reference  &Tuning range (GHz) & Linewidth (MHz) & Output power (dBm) & Applied bias field & TMR/$R_\text{P}$($\Omega$) & Size     \\ \hline 
MTJ NP&\cite{Nazarov2008a}, \scriptsize{2008}  &4--7* & 21 & -47 & In-plane & 48\%/ 16.5  & 250 nm circular   \\ \hline 
MTJ NP&\cite{Villard2010a, Houssameddine2008a}, \scriptsize{2010}  &4--10** & 26 & -46** & In-plane & 100\%/ 275**  & 80 nm circular     \\ \hline 
MTJ NP&\cite{Muduli2011a, Mufuli2009}, \scriptsize{2011}  &3--12** & 20** & -38.5 & In-plane & 70\%/ 42.5  & 240 nm circular     \\ \hline 
MTJ NP***&\cite{Kubota2013}, \scriptsize{2013}  & 4--7** & 47 & -32.6  & Out-of-plane & 66\%/ 180** & 120 nm circular\\ \hline 
MTJ sombrero-shaped NC &\cite{Maehara2013}, \scriptsize{2013}  & 2--5**** & 12 & -26.2 & In-plane & 46\%/ -- & 50$\times$150 nm$^2$ \\ \hline 
MTJ sombrero-shaped NC &\cite{Maehara2014}, \scriptsize{2014}  & 2.5--15** & 3.4 & -42 & Out-of-plane & 38\%/
 55      & 50$\times$150 nm$^2$ \\ \hline 
MTJ vortex (10 nm FeB) &\cite{Tsunegi2014}, \scriptsize{2014}  & 0.2--0.5**** & 0.074 & -28.5 & Out-of-plane & 128.5\%/ 35**      & 300 nm circular \\ \hline
Single-layer-NiMnSb based NC &\cite{Durrenfeld2015b}, \scriptsize{2015}  & 1--5 & 0.5 & -105 & In-plane & -- /
 22.4      & 90 nm circular \\ \hline 
GMR NC &\cite{Rippard2014}, \scriptsize{2004}  &9.7--34.4** & 1.89 & -74** & Out-of-plane & 1\%/ $\sim$15  & 40 nm circular   \\ \hline 
GMR NP &\cite{Mistral2006}, \scriptsize{2006}  &11.4--12.25** & 3.2 & -77** & In-plane & 0.44\%/ 13.62  & 50$\times$100 nm$^2$   \\ \hline 
GMR NC &\cite{Bonetti2009}, \scriptsize{2009}  &10--46** & 4.5 & -- & Out-of-plane & -- / --  & 40 nm circular   \\ \hline 
GMR NP &\cite{Sinha2011}, \scriptsize{2011}  &10--11**** & 10 & -53 & In-plane & 2.75\%/ 8.72  & 130$\times$90 nm$^2$  \\ \hline 
GMR NP &\cite{Seki2013a}, \scriptsize{2013}  &$\sim$3.5--4**** & --& -60 & In-plane & 12\%/ 8.4**  & 110$\times$60 nm$^2$  \\ \hline 
GMR NP &\cite{Seki2014a}, \scriptsize{2014}  &$\sim$4--14** & 10 & -46***** & Out-of-plane & 48\%/ 5.88 & 170$\times$100 nm$^2$  \\ \hline 
GMR NC &\cite{Yamamoto2015a}, \scriptsize{2015}  &$\sim$0.8--2 & 3 & -46 & Zero-field & 10\%/ 9.55** & 140 nm circular  \\ \hline 
\hline
\end{tabular}
\begin{tablenotes}
      \scriptsize
      \item NP, NC: nanopillar, nanocontact.
      \item * The frequency range in which narrow peaks are observed.
      \item ** Data extracted from figures.
      \item ***  MTJ STNO has perpendicularly magnetized FL and in-plane magnetized PL.
      \item **** full tuning range not provided.
      \item ***** Maximum RF output power after de-embedding is -35 dBm. 
\end{tablenotes}
\end{threeparttable}
\end{table*}
\end{center}

To highlight the advantages of STNOs, we compare them with the common types of microwave oscillators, including voltage-controlled oscillators (VCOs), yttrium iron garnet (YIG) tuned oscillators (YTOs), and dielectric resonator oscillators (DROs) \cite{Khanna2015}, as summarized in Table II. Among the listed microwave oscillators, STNOs have the smallest size and the fastest switching speed. STNOs also surpass VCOs and DROs in tunability range and outperform YTOs in power consumption. 

\begin{center}
\begin{table*}[tb]
\centering
 \caption{Comparison between STNOs and common types of microwave oscillators}
\label{table:kysymys}
\small
  \centering
  \begin{tabular}{| p{1.5cm} | p{2.2cm} | p{3.2cm}| p{2.2cm} | p{2.25cm} | p{1.5cm}| p{2.2cm} |}
\hline\hline
Type  &Tuning range& 10 GHz phase noise @100 kHz offset
 & Switching speed & Size (inch cube)  & Cost&  Power consumption      \\ \hline
  VCO \cite{Khanna2015}  & Octave & -110 dBc/Hz
 & 1 $\mu$s & 0.001   & Low &   Low     \\ \hline
  YTO \cite{Khanna2015} & Decade & -120 dBc/Hz
 & 1 ms & 1  & High &   High     \\ \hline
  DRO \cite{Khanna2015}  & 1 \% & -120 dBc/Hz
 &  N/A & 0.5 & Medium  &   Low     \\ \hline
  \textbf{STNO}  & Multi-octave &  N/A*
& 1 ns &  $1\times10^{-16}$ &   Low &   Low   
 \\\hline \hline
\end{tabular}
\begin{tablenotes}
      \scriptsize
      \item *  The phase noise of the MTJ STNO was  measured by Nanosc AB at about -65 dBc/Hz at 1 MHz offset.
% cannot find the reference since the webpage was delete    
\end{tablenotes}
\end{table*}
\end{center}

The issues of spectrum impurity and the limited output power must be overcome before STNOs enter  widespread use as microwave sources. Recently, significant progress has been achieved in understanding the fundamentals and in overcoming these two limitations of STNOs \cite{ Maehara2013,Maehara2014, pogoryelov2011a, Quinsat2010a, Eklund2014, Seki2013a, Seki2014a, Yamamoto2015a}. As  can be seen in Table I, a sombrero-shaped nanocontact geometry for MTJs has been developed \cite{Maehara2013, Maehara2014} that remedies the incompatibility of conventional nanocontact geometry with MTJs. This geometry enables the optimization of either the Q factor (narrower linewidth) \cite{Maehara2014} or the output power \cite{Maehara2013}. Specifically, a very narrow linewidth of 3.4 MHz \cite{Maehara2014} and an output power of -26.2 dBm \cite{Maehara2013} were obtained. These newly developed MTJ STNOs show significant improvement in either linewidth or output power in comparison with the conventional MTJ STNOs \cite{Nazarov2008a, Mufuli2009, Kubota2013}. In addition, recent research has shown that by using a tilted CoFeB FL \cite{Skowronski2012}, the bias field can be removed, which opens up more possibilities for MTJ STNO-based applications. 
For GMR STNOs, which typically have a narrower linewidth than MTJ STNOs, the output power has been enhanced by using CoFeMnSi (CFMS) layers as magnetic layers \cite{Seki2014a}. By further combining the CFMS with nanocontact geometry, a large output power of -46 dBm and a narrow linewidth of 3 MHz can be achieved simultaneously at zero applied field \cite{Yamamoto2015a}. 
In addition, the phase noise of STNOs has been reduced by using mutual synchronization and current modulation as an extrinsic tool \cite{Sani2013a, pogoryelov2011a, Dumas2014a}. The limited output power has been improved by employing amplifiers dedicated to STNOs \cite{Villard2010a, Chen2014a}. The achieved improvements of phase noise and output power bring STNOs closer to being used as microwave sources. 

In addition to these issues, the device-to-device performance variation of STNOs is significant. Furthermore, the mode structure and temperature dependence of STNOs need to be well understood so as to suppress the unwanted modes and to estimate the performance variation that results from changes in temperature. Other issues that need to be taken into consideration in applications include the irreversible changes of STNOs and thus reliability issues at high biasing current \cite{Houshang2014a}. 

\begin{figure}[tb]
  \begin{center}
    \includegraphics[trim = 9mm 7mm 11mm 5mm, clip, width=8.87cm]{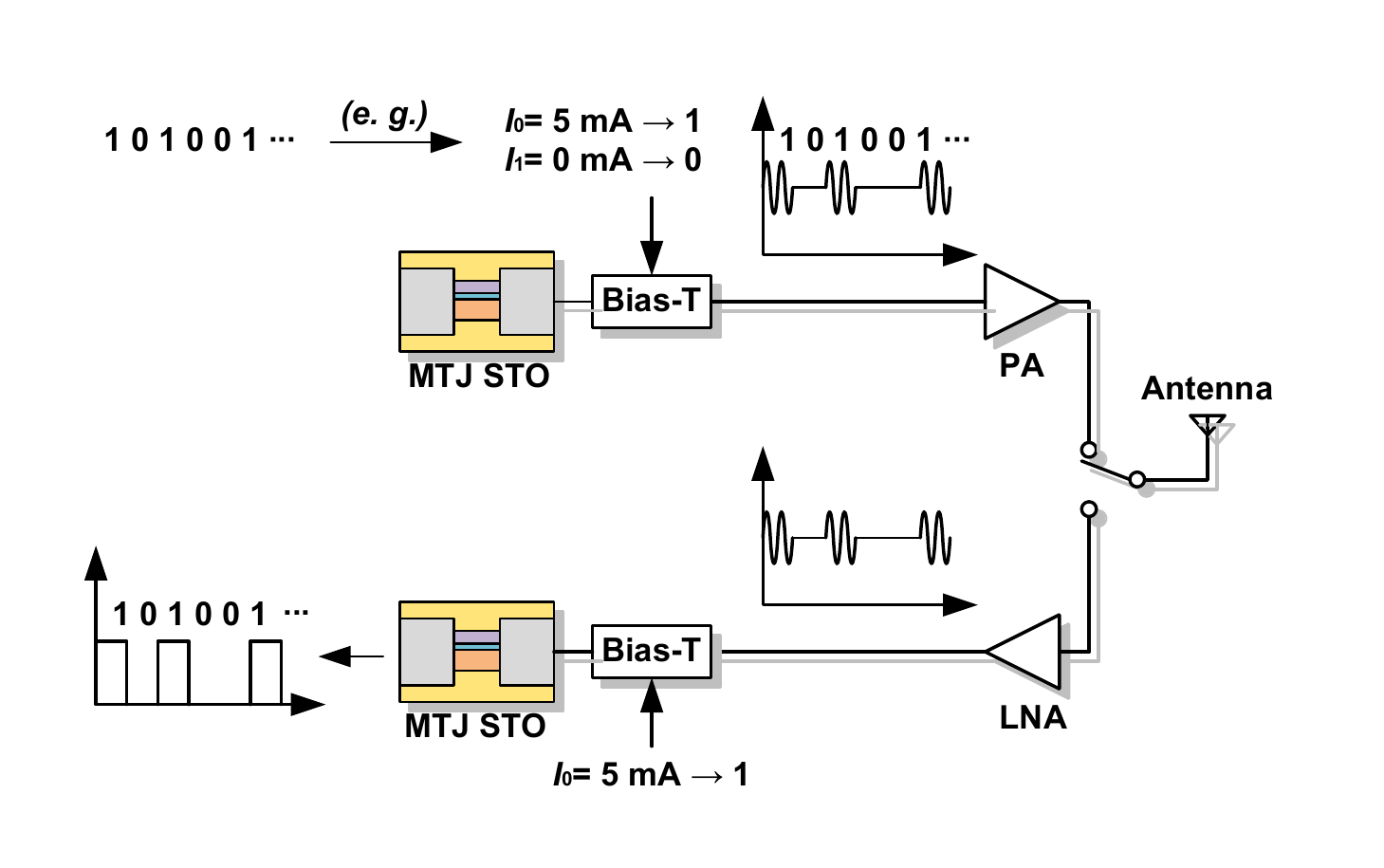}
    \caption{System architecture of an MTJ STNO-based transceiver}
    \label{STO_based_transceiver}
  \end{center}
\end{figure}

\subsubsection{Microwave detectors}
The STT provides many useful phenomena as detailed in \cite{Locatelli2014a}. One important phenomenon is the spin-torque diode effect discovered in  MTJ-based devices \cite{Tulapurkar2005a}. The spin-torque diode effect can be described as follows: when a microwave signal with a frequency close to the natural FMR frequency of one of the MTJ's electrodes is incident on the device, the induced oscillating tunnel current  can excite magnetic precession via spin transfer. The resistance oscillation that results from this precession is combined with the oscillating current so as to produce a measurable DC voltage component across the tunnel junction \cite{Tulapurkar2005a, Ishibashi2010a, Wang2009a}. Recently, an MTJ microwave detector that does not require any external bias fields, and that possesses  a  sensitivity of about 75400 mV/mW has been demonstrated at room temperature \cite{Fang2014}. In addition, this microwave detector can provide very high sensitivity even at a low input microwave powers of 10 nW (-50 dBm). Its sensitivity is about one order of magnitude higher than that of a semiconductor diode, such as a Schottky diode. 
The large sensitivity of MTJ STNOs makes them promising candidates as frequency-tunable resonant microwave detectors for a wide range of applications in telecommunications, radars, and smart networks \cite{Fang2014}. 

\subsubsection{Noncoherent transceiver}
Taking advantage of the fast switching speed of STNOs, an STNO-based wireless communication system with binary amplitude shift keying (ASK) has been demonstrated in \cite{Choi2014a}. In this system, the STNO performs the ASK modulation at the transmitter and a simple envelope detector recovers the data at the receiver. For this noncoherent communication system, the considerable phase noise of STNOs is not an issue. An STNO with 1 ns turn-on time (or switching speed) allows a communication system that can achieve a data rate as high as 1.48 Gbps \cite{Krivorotov2005a, Choi2014a}.

Considering also the spin-torque diode effect, STNOs can be utilized as both modulators and demodulators in transceiver systems, as shown in Fig. \ref{STO_based_transceiver}.  As modulators, STNOs translate different currents representing the binary data ``0'' and ``1'' directly into signals at  different frequencies. As demodulators, STNOs enable the direct conversion of the received signals to high and low output voltages representing ``0'' and ``1''.  These STNO-based transceivers may enable the development of low-complexity, low-cost, high data-rate, and noncoherent communication systems.

\subsubsection{Magnonics}
STNOs with a point contact (Fig. 2(a)) on an extended free layer allow spin-torque-driven emission as well as propagation of spin waves outward from the nanocontact \cite{Locatelli2014a, Madami2011a}. The spin wave emitted from an STNO with a nanoscale electrical contact has been directly observed and can propagate for several micrometers \cite{Madami2011a}. The emission of propagating spin waves is the basis of synchronizing arrays of STNOs for microwave applications. Moreover, STNOs are very attractive as wave emitters for future use in magnonic devices. 

Magnonics is an emerging and rapidly growing research field that connects magnetism, spintronics, and electronics \cite{Bonetti2013a}. It investigates and utilizes magnetic phenomena connected with spin waves, in a broad sense. Magnonics employs different devices to deal with spin waves. The key devices are spin-wave emitters, spin-wave manipulators (modulators), and spin-wave detectors. STNOs, as previously discussed, can play all the roles of these key devices. As a consequence, STNOs have the potential to realize all the fundamental functions that are necessary for magnonics \cite{Locatelli2014a, Bonetti2013a}. 

\subsubsection{Neuromorphic computing} 
Neurons can be modelled as nonlinear oscillators that vary their rhythms based on incoming signals \cite{Locatelli2014a, Izhikevich1999, Izhikevich2004}. To form a network of neurons coupled by synapses, the nonlinear oscillators need to be able to synchronize with each other in frequency or phase. The neural network can therefore be realized using the non-Boolean nature of these nonlinear oscillators \cite{Zhao2015, Sharad2015}. STNOs are able to act as nonlinear oscillators because they can  synchronize when coupled via a mutual electrical or magnetic interaction. Furthermore, it has been demonstrated that other spintronic devices---including spin-torque memristors and spin-torque memories---are very suitable in playing the roles of synapses and memory processes in mimicking neural networks \cite{Locatelli2014a}. Consequently, STNOs, together with other spintronic devices, pave the way to full spintronic implementations of neural networks, including both polychronous wave computation and information storage \cite{Sharad2012}. 

Non-Boolean computing based on STNOs shows great potential in data-processing applications, particularly in neural networks. STNO-based non-Boolean computing is discussed in \cite{Sharad2015, Roy2014a} and its physical processes are reviewed in \cite{Pufall2015a}. Particularly, the concept of a dual-pillar STNO that decouples bias and read paths for non-Boolean energy-efficient computing has been proposed, and a simulation framework has been presented in \cite{Roy2014a}. The dual-pillar STNO is composed of a GMR STNO and an MTJ STNO. The GMR STNO is employed for the bias path so as to achieve a low-resistance interface and hence to allow low bias-voltage, as well as high energy efficiency. The high-resistance MTJ STNO is used to sense a small current and then to  produce an output signal with large amplitude.

\subsubsection{Magnetic field sensing}
The magnetoresistive effect, which is necessary in the operation of STNOs, is also widely used to translate detected magnetic field strength to resistance. Specifically, GMR-based sensors have been commercialized for sensing magnetic fields which range in strength from the Earth's field up to a couple of hundred Oersted \cite{Caruso1998}. In addition to this range, larger magnetic fields can also be measured by STNOs, by virtue of measuring the change in their oscillation frequency caused by the difference between the bias field and the magnetic field being tested. This STNO-based magnetic field sensor can provide high sensitivity up to 180 GHz/T \cite{Braganca2010a}. Moreover, the output of the STNO is the frequency or frequency change, rather than the signal amplitude, thus relaxing the requirement for the system to have a large signal-to-noise ratio. For this reason, STNOs are very attractive for sensing large magnetic fields.

\subsubsection{Magnetic field generation}
The concept of employing STNOs to generate localized AC magnetic fields for microwave-assisted magnetic recording (MAMR) applications has been proposed in \cite{Zhu2008a}, theoretically analyzed in \cite{Zhang2015}, and extensively studied in \cite{Okamoto2015}. Toshiba recently demonstrated a new technology that uses the localized AC magnetic field to reverse the magnetization direction,  allowing the excitement of a magnetization oscillation in only one specific magnetic layer of a multilayer magnetic recording system \cite{Toshiba2015}. This technology could enable the development of 3-D magnetic recording systems, where independent data are written and read from overlapping layers of a multilayer recording medium.

The applications to which STNOs are suited are not limited to the above suggestions. For instance, STNOs have been found very useful as high-density massively parallel microwave signal processors and small phased array transmitters \cite{Sattler2010a}. It has also been demonstrated in \cite{Sato2012a} that STNOs can be employed to develop novel readers with high data transfer rates for hard disk drives (HDDs). Other applications that would make full use of the nanoscale aspects and of the high-speed broad tunability of STNOs need  further exploration.

\subsection{SHNO-based applications: state-of-the-art and prospects}
With the recent emergence of the spin Hall effect, a new type of spintronic oscillator, the SHNO, has been developed \cite{Demidov2012}. This provides a new route for the development of microwave and magnonic devices  \cite{Durrenfeld2015a}. SHNOs exhibit several advantages over STNOs,  including easier nanofabrication, lower required DC current, direct optical access to the magnetodynamically active area, smaller radiation losses, and suppressed nonlinear damping process \cite{Ranjbar2014, Durrenfeld2015a}. However, the linewidth of SHNOs is on the same order as that of STNOs, and their output power is currently lower than that of STNOs. Despite the issues with SHNOs, their advantages offer an opportunity to implement novel nanoscale microwave sources and emitters for wireless communications and magnonics applications \cite{Demidov2015a, Giordano2014a}.

\subsection{STNO/SHNO-based applications: issues, challenges, and developments} 
In addition to the need to improve STNO and SHNO performance, STNO electrical models, STNO and SHNO integration into integrated circuits (IC), and dedicated circuitry are also necessary for the development of applications based on STNOs and SHNOs. Electrical models of STNOs and SHNOs are critical, since they can efficiently reduce the time, cost, and risk of errors involved in building prototypes of the entire system of the STNO or SHNO with the IC. Integration of the STNO or SHNO with the IC is important, since it should enable better performance, miniaturization, and even single-chip solutions.
Dedicated circuitry is essential for STNO/SHNO-based applications, since it can perform necessary functions such as biasing and amplification, and can extrinsically improve the performance of STNOs or SHNOs.
The issues, challenges, and development of the electrical models, integration, and dedicated circuitry are discussed in the following.

\subsubsection{Electrical models}
The basic procedure for  modeling STNOs or SHNOs involves solving the Landau-Lifshitz-Gilbert equation with a Slonczewski spin-transfer term (LLGS). One possible approach to solving the LLGS is based on micromagnetics \cite{Firastrau2008}. Such a model presents the magnetization dynamics of STNOs and SHNOs on a microscopic scale. It typically considers the anisotropy energy, the Zeeman energy of the external field, the ferromagnetic exchange energy and the magnetostatic energy \cite{fidler2000, Berkov2008}. 
However, the micromagnetic simulation typically takes days, which is not efficient for the development of STNO/SHNO-based applications. Additionally, the output dataset of this model cannot be directly used in a circuit simulator. The complicated numeric computation involved in the micromagnetics-based model prevents it from being implemented in a hardware description language. A hardware description language is a specialized computer language used to describe both structural and behavioral aspects of designs including, but not limited to, electronic components, circuits, and systems \cite{vachoux2012}. It allows the automatic analysis, simulation, and evaluation of designs, through the use of electronic design automation (EDA) tools. The micromagnetics-based model is therefore not suited to evaluating  STNOs and SHNOs under different bias conditions together with other circuits, as would be required in STNO/SHNO-based applications. 

An alternative approach to solving the LLGS is based on the macrospin approximation, which assumes that only a spatially uniform magnetization precession is excited and that the spin-polarized current is uniform across the area of the free layer \cite{Slavin2009a}. A macrospin-based model is able to provide an analytical solution with acceptable accuracy, while allowing implementation in a hardware description language. As a result, a macrospin-based analytical model is needed to design and develop STNO-based systems. Several macrospin-based models of STNOs \cite{Chen2015a, Chen2015b, Lim2013a, Kim2013a, Ahn2013a} have been developed and implemented in Verilog-A, a commonly used hardware description language for analog systems. However, the models in \cite{Lim2013a, Kim2013a, Ahn2013a} require the aid of Matlab to compute the effective magnetic field, which is then manually imported into the Verilog-A model. Consequently, they cannot be considered fully implemented in Verilog-A. 
Moreover, these models \cite{Lim2013a, Kim2013a, Ahn2013a} use equations that are not fully validated. Additionally, only one specific device has been used to verify the models, which is not sufficient. A comprehensive MTJ STNO model that can overcome the issues in \cite{Lim2013a, Kim2013a, Ahn2013a}, has been presented in \cite{Chen2015a, Chen2015b}. This further allows accurate replication of the phase noise and hence of the generated signal. Moreover, this model can be used for various MTJ STNOs to estimate their performance together with the ICs. Nonetheless, this model is not suitable for simulations that involve momentary variations in the operating conditions, such as current or field modulation. To enable comprehensive analysis and evaluation of STNOs or SHNOs together with ICs, future models need to be divergence-free, while allowing all the necessary simulations and analyses. Future models also need to include temperature dependence.

\subsubsection{STNO/SHNO-IC integration}
To avoid the losses due to external components and connections, as well as to eliminate wave reflections, STNOs and SHNOs need to be integrated with other technologies, such as silicon-CMOS and gallium-arsenide (GaAs), which allow the circuits required for STNOs and SHNOs to be implemented. Additionally, integration is vital if the nanoscale of STNOs and SHNOs is to be capitalized upon and to achieve compact STNO/SHNO-based systems. 
The integration approach must be considered before implementing the circuits and developing the system, since it has a significant impact on the system topology and circuit design. The advantages and disadvantages of different possible integration approaches are discussed in \cite{Chen2015c}. The wire-bonding  approach has been successfully used in \cite{Villard2010a, Chen2015c}, mainly thanks to its independent preintegration characterization and optimization of the IC and STNO. In \cite{Villard2010a}, an MTJ STNO has been integrated with a dedicated CMOS IC amplifier and an off-chip bias-tee using wire bonding. However, the use of the off-chip bias-tee limits both the integration level and the bandwidth of the system. The CMOS IC proposed in \cite{Chen2015c} includes an on-chip bias-tee and  has been successfully integrated with an STNO by bonding wires. The bonding wire has also been used as part of the bias-tee design, since it introduces extra high-frequency impedance into the AC decoupling path of the bias-tee. 
%Using the bonding wire in the bias-tee design demonstrates that the integration approach can affect the circuit design.
Furthermore, the parasitics of the bonding wires in the signal path, as well as the on-chip bias-tee, determine the requirements of the input and output filters for achieving a flat frequency response over a large bandwidth. The wire-bonding approach is appropriate for the current level of maturity of STNO technology, as it allows easy implementation, measurement, and replacement. 
\begin{figure}[tb]
  \begin{center}
    \includegraphics[trim = 30mm 30mm 24mm 14mm, clip, width=9cm]{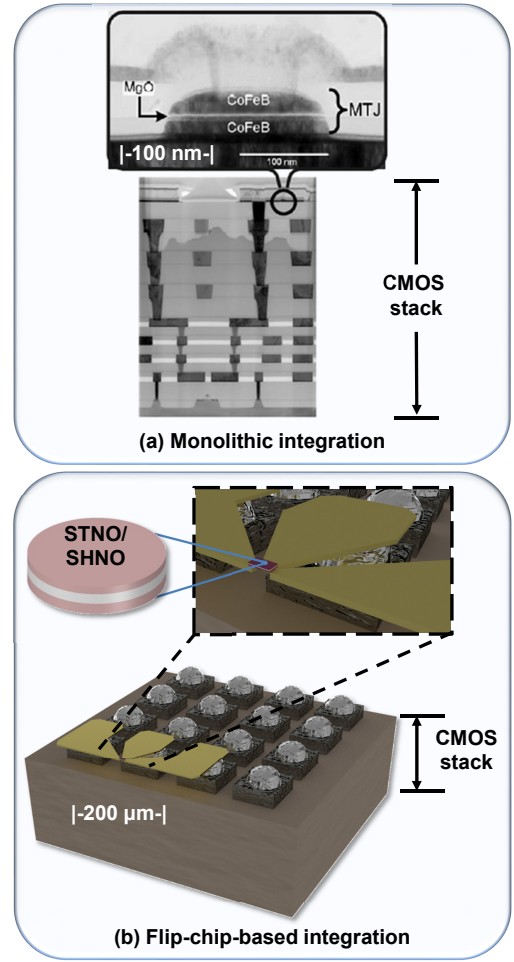} 
    \caption{(a) Monolithic integration (with 200 nm length scale) \cite{Fujitsu2010}, and (b) flip-chip-based integration approaches (with 200 $\mu$m length scale)}
    \label{STO_IC_integration}
  \end{center}
\end{figure}

Monolithic and flip-chip-based approaches can ultimately result in a higher level of integration and better performance at high frequencies. The monolithic integration approach has been successfully used in hybrid CMOS/MTJ logic circuits (Fig. \ref{STO_IC_integration}(a)) targeting the ST-MRAM technology \cite{Fujitsu2010, ikegami2014, Rizzo2013}, which utilizes the same STT effect as that in STNOs and SHNOs. For instance, a fully functional 64 Mb ST-MRAM built on a standard 90 nm process has been developed in  \cite{Rizzo2013}. Nevertheless,  monolithic integration has not yet been implemented so as to handle the RF signals generated by STNOs or SHNOs. In order to use monolithic integration, the top-layer surface roughness of the IC stacks needs to be very well controlled, as  this layer acts as the Cu buffer layer for fabricating STNOs and SHNOs. The roughness of this layer has significant impact on the spatial variation in the multilayer structure, and hence on the dynamics of STNOs and SHNOs \cite{Shaw2010}. An RMS roughness lower than $\sim 3$ {\AA} is necessary to avoid unpredictable device performance and significant performance degradation \cite{Shaw2010}. Flip-chip-based integration is one of the principal methods of interconnecting ICs \cite{Elenius2000}. As the cost of flip-chip-based integration decreases \cite{Pang2015}, its application space  broadens further. Fig. \ref{STO_IC_integration}(b) shows a possible integration between an  STNO/SHNO and a CMOS IC using the flip-chip-based approach. Compared to  monolithic integration,  flip-chip-based integration allows easier implementation and measurement and retains a low-loss interconnection \cite{Chen2015c}.
However, neither monolithic nor flip-chip-based integration will be suitable until the device-to-device performance has improved to the extent that no major adjustments (such as to the topology) of the circuitry are required for different devices.

\subsubsection{Dedicated circuitry for STNOs/SHNOs}
The STNO/SHNO-based applications described above could be possibly achieved by integrating STNOs or SHNOs into the existing system architectures and by employing the existing circuits implemented in other technologies \cite{TamaruINTERMAG2015, Villard2010a, Chen2014a, Choi2014a, Sattler2010a}. %% probably more?
For instance, a system architecture similar to the simple noncoherent transceiver in \cite{Byeon2011} has been used to develop an STNO-based wireless communication system \cite{Choi2014a}. An amplifier using the existing topology of a cascode differential pair with a resistive load has been implemented in \cite{Villard2010a} to provide amplification for the STNO. 
Additionally, considering the device-to-device performance variation of STNOs and SHNOs, dedicated circuitry may need to allow adjustable performance so as to avoid multiple designs for different STNOs or SHNOs.
For instance, the variable gain amplifier (VGA) \cite{Wagner2014} and the tunable LNA \cite{Chen2005}, whose gain and bandwidth, respectively, can be tuned  to some extent, may be suitable for providing adjustable gain and bandwidth for different STNOs or SHNOs. 

However, several dedicated circuits need to be designed for STNOs and SHNOs. The first important circuit is the bias-tee, which is necessary for applying the required DC bias current into STNOs and SHNOs without disturbing the RF signals generated. Current mirrors have been considered in \cite{Lim2013a, Kim2013a} for biasing STNOs. However, the resistance (biasing voltage) of the STNO changes as the angle of the magnetic field is varied, so the current mirror cannot accurately copy the current from a current source to the STNO under all circumstances. Accordingly, an on-chip bias-tee that can provide accurate current for STNOs and SHNOs is necessary. Factors that also need to be taken into account during the design of the bias-tee include the maximum applied DC current, the applied integration approach, the resistance of STNOs and SHNOs, the power supply voltage of IC technology, the clamp circuit for providing ESD protection, and the parasitics. A dedicated sample bias-tee design is given in \cite{Chen2015c}. This design considers the above factors and takes care of the trade-offs involved in selecting the number of turns, the separation distance between each turn, and the width of metalization needed for the inductor to act as the AC decoupling component. To implement the dedicated on-chip bias-tee, electromagnetic simulations may be required when custom inductors are necessary \cite{Bell2003, Bell2007}.

The second important circuit for STNOs and SHNOs is the amplifier, which needs to provide sufficient gain and bandwidth for a given application while featuring low noise and proper input impedance. For example, to use STNOs or SHNOs as LOs, the amplifier should typically enhance the output power of the STNOs and SHNOs to 0 dBm so as to drive an RF mixer to perform frequency conversion. %An MTJ STNO with -26 dBm output power \cite{Maehara2013} then requires a low noise amplifier (LNA) with a gain larger than 26 dB. 
To take full advantage of STNOs or SHNOs in multiband multistandard radios, the bandwidth of the amplifier needs to cover multiple bands of interest. Additionally, the amplifier must introduce as little noise as possible to the entire system. The input impedance of the amplifier needs to match the impedance of STNOs or SHNOs for maximizing the transferred power. Nevertheless, for STNOs or SHNOs with a small resistance (a few Ohms), wideband impedance matching to such a resistance is not feasible. This is because any parasitic will introduce an impedance mismatch much larger than a few Ohms. In this case, rather than maximizing the transferred power, an amplifier with large input impedance is preferable, since it can maximize the AC voltage delivered to the amplifier as well as the signal-to-noise ratio at the output \cite{Chen2015c}.

Other important circuits include those that handle the phase noise of STNOs and SHNOs, such as phase-locked loops (PLLs) and injection-lock ring oscillators (ILROs) for frequency synthesis. This is due to the fact that the frequency fluctuation of STNOs occurs every several nanoseconds or even few tens of nanoseconds \cite{Villard2010a, Chen2015a}. The fast frequency change requires a rapid lock time (or settling time) of at most a few nanoseconds, which is difficult to achieve in either PLLs or ILROs. Recently, the phase of an MTJ STNO, which generates a 5.12 GHz microwave signal, has been successfully locked by a PLL to that of an 80 MHz reference clock, resulting in a significant increase in Q from 800 to 5$\times 10^9$ \cite{TamaruINTERMAG2015}. This is the first 
demonstration of  phase synchronization of STNO output to a lower frequency external signal, and it may be a milestone toward practical applications of STNOs.

\section{Conclusion}

State-of-the-art STNOs and SHNOs feature very wide tunability, high levels of integration with CMOS, high operating frequencies, fast switching speeds, and miniature size. Their unique blend of features suggests a wide range of applications, including microwave sources, microwave detectors, noncoherent transceivers, magnonics, neuromorphic computing, magnetic field sensing, and magnetic field generation. Over the past decade, extensive research has taken place to understand and further exploit the nonlinear physical phenomena of STNOs and SHNOs.
In this paper, we have introduced the fundamentals, geometries, and functional properties of STNOs and SHNOs. We have additionally reviewed the state of the art in STNO and SHNO technologies. Furthermore, we have presented the potential for these oscillators in their popular applications. 
However, challenges still remain on both the device and system level. These need to be overcome before the use of STNOs and SHNOs can become widespread. %%%re-write

% or


%\appendices
%Some text for the appendix.

\section*{Acknowledgment}

%%%%Is this right?
This work is partially supported by the Swedish Research Council (VR), %under Grant 2009-4190,  
the Swedish Foundation for Strategic Research (SSF), the Knut and Alice Wallenberg Foundation (KAW), the European Commission FP7-ICT-2011-contract No. 317950 ``MOSAIC'', and the European Research Council (ERC) under the European Community's Seventh Framework Programme (FP/2007-2013)/ERC Grant 307144 ``MUSTANG''.%Johan {\AA}kerman is a Royal Swedish Academy of Sciences Research Fellow supported by a grant from KAW.

\ifCLASSOPTIONcaptionsoff
  \newpage
\fi



%\bibliographystyle{IEEEtran}

%\begin{thebibliography}{1}
%\end{thebibliography}

%\bibliographystyle{unsrt}      %This compiles author names correct, but not vol and pp no.
\bibliographystyle{IEEEtran}    %This compiles vol and pp no correct, but not all authors.

%\begin{biography}[{\includegraphics[width=1in,height=1.25in,clip,keepaspectratio]{mshell}}]{Michael Shell}

\begin{IEEEbiography}[{\includegraphics[trim = 30mm 62mm 45mm 45mm  width=1in,height=1.25in,clip,keepaspectratio]{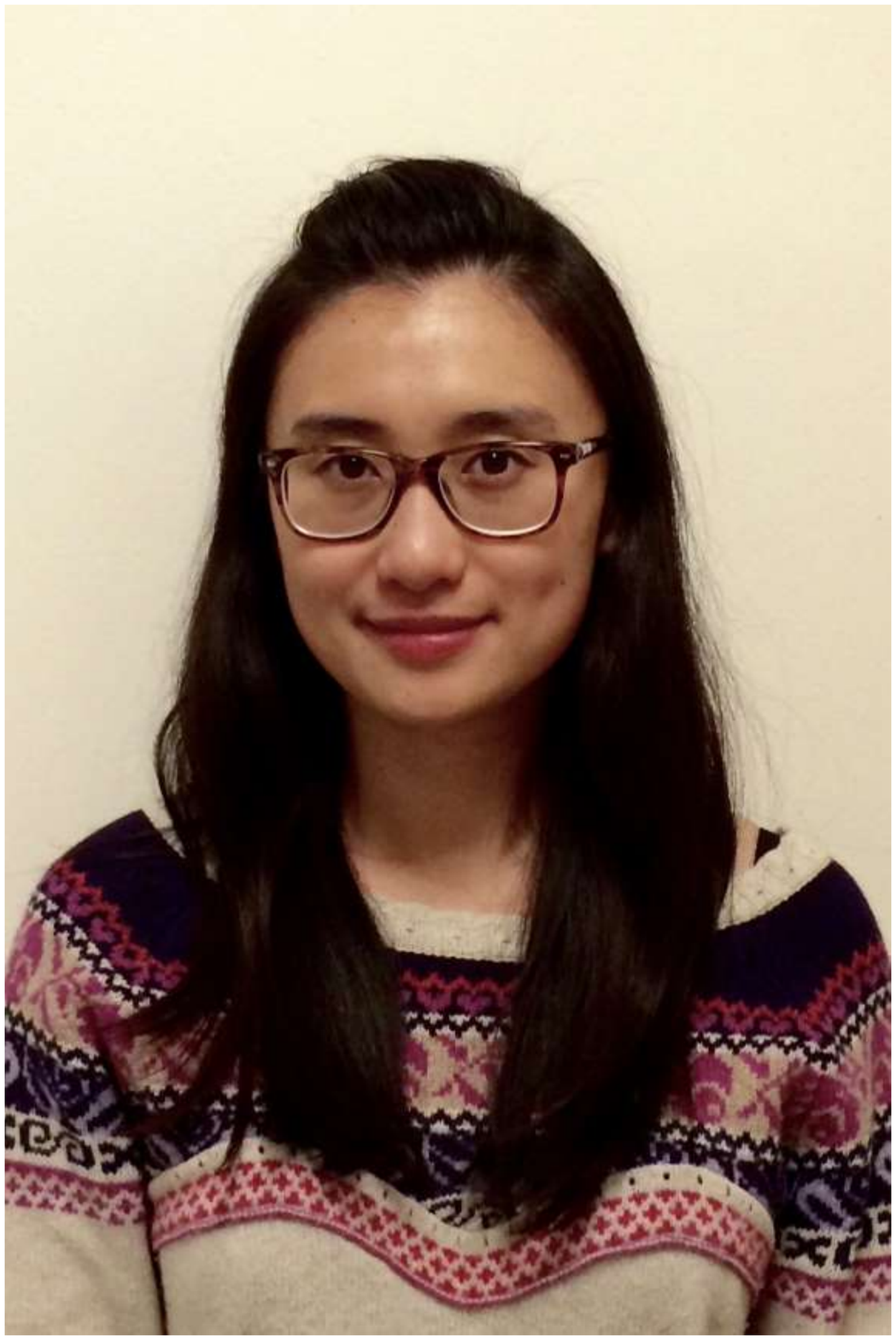}}]
{Tingsu Chen}
(S'11) received an M.Sc. degree in system-on-chip design from the KTH Royal Institute of Technology, Sweden, 2011. She is currently working toward a Ph.D. degree at KTH in the research area of modeling spin-torque oscillators and designing high-frequency circuits for spin-torque oscillator technology.
\end{IEEEbiography}

\begin{IEEEbiography}[{\includegraphics[trim = 17mm 24mm 115mm 22mm  width=1in,height=1.25in,clip,keepaspectratio]{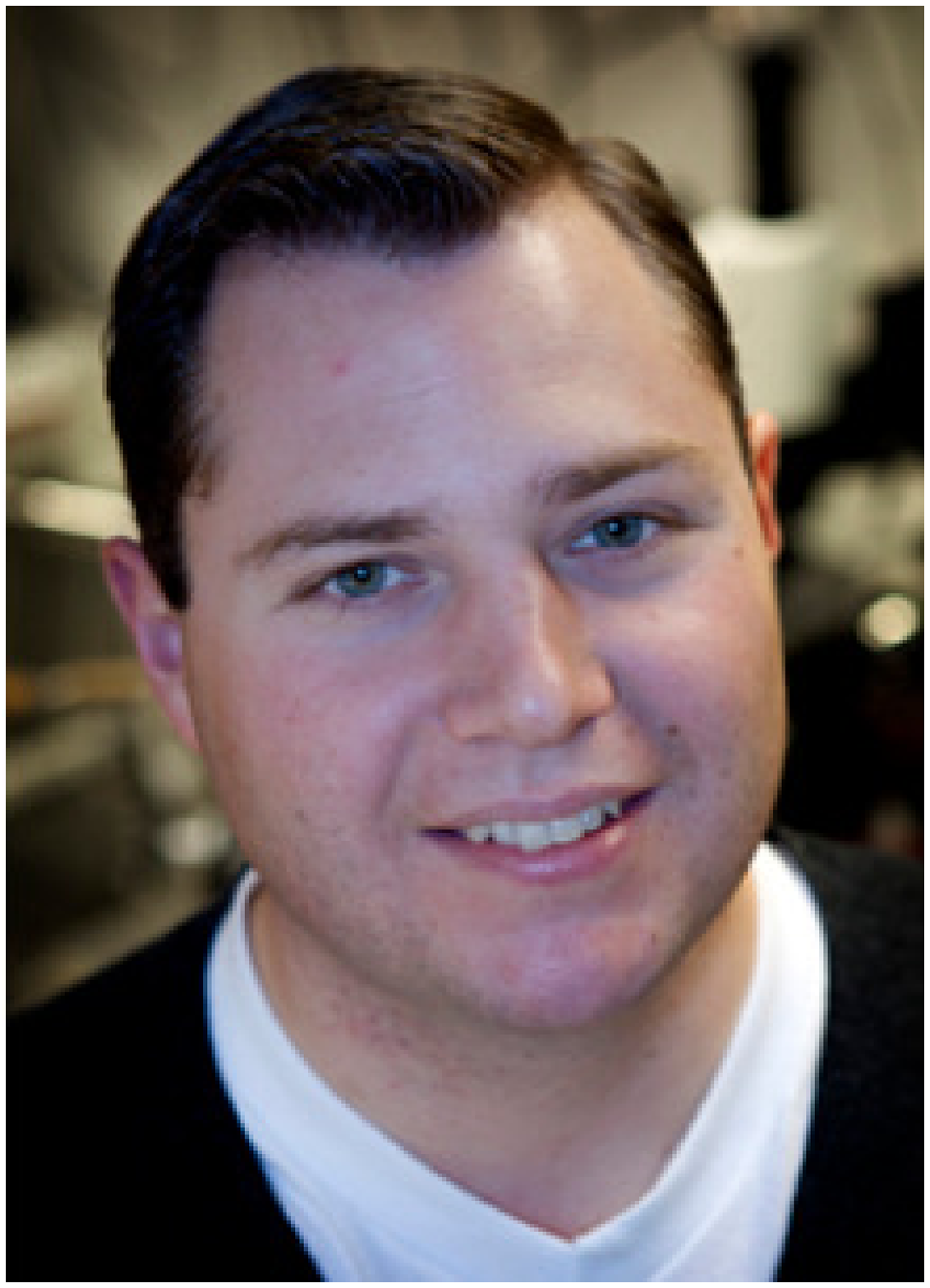}}]{Randy K. Dumas}
(Member IEEE) received his Ph.D. in Physics from the University of California, Davis on the topic of reversal mechanisms in magnetic nanostructures in 2009.  From 2009--2012, he was a postdoctoral research fellow at the University of Gothenburg, where he is currently a permanent researcher.  His research interests include the fabrication and characterization of magnetic nanostructures with a current focus on magnonics. 
\end{IEEEbiography}

\begin{IEEEbiography}[{\includegraphics[trim = 25mm 117mm 72mm 40mm width=1in,height=1.25in,clip,keepaspectratio]{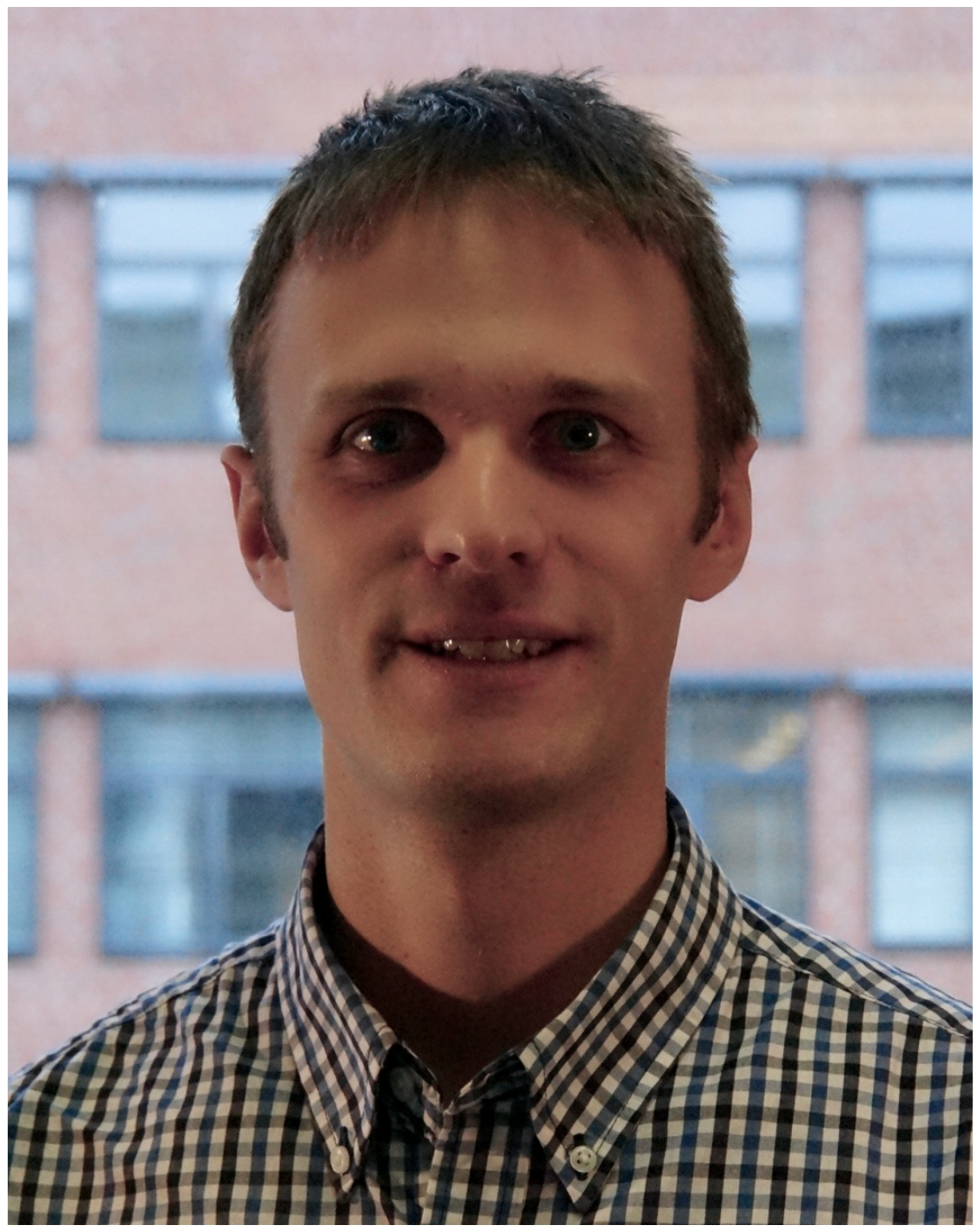}}]
{Anders Eklund}
(S'13) received an M.Sc. degree in engineering physics from KTH Royal Institute of Technology, Sweden, in 2011. He is currently working towards a Ph.D. degree in physics at KTH, experimentally investigating the frequency stability and mode structure of spin-torque oscillators by means of electrical characterization and synchrotron x-ray measurements.
\end{IEEEbiography}

\begin{IEEEbiography}[{\includegraphics[width=1in,height=1.25in,clip,keepaspectratio]{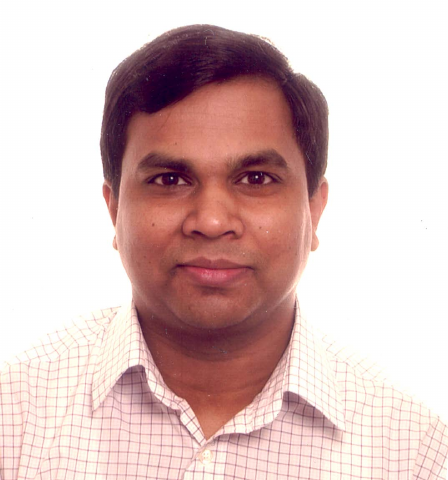}}]{Pranaba K. Muduli}
(Member IEEE) received his Ph.D. degree in physics from the Humboldt University, Berlin, Germany in 2005. He did postdoctoral research work at the University of North Carolina, USA, the Royal Institute of Technology, Sweden, and the University of Gothenburg, Sweden. Since July 2012, he has been an Assistant Professor at the Indian Institute of Technology Delhi, India. He is also a researcher at the University of Gothenburg, Sweden. His current research interests include spintronics and nanomagnetism with an emphasis on spin-torque induced magnetization dynamics. 
\end{IEEEbiography}

\begin{IEEEbiography}[{\includegraphics[width=1in,height=1.25in,clip,keepaspectratio]{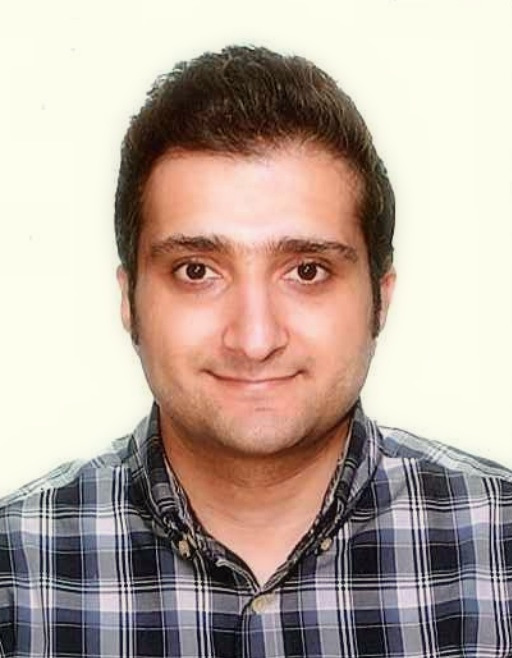}}]
{Afshin Houshang}
(S'12) received his M.Sc. in electroceramics engineering from Shiraz University, Shiraz, Iran, 2010.  He is currently working towards a Ph.D. degree in physics at the University of Gothenburg in the area of synchronization phenomena in spin-torque and spin Hall oscillators. 
\end{IEEEbiography}

\begin{IEEEbiography}[{\includegraphics[width=1in,height=1.25in,clip,keepaspectratio]{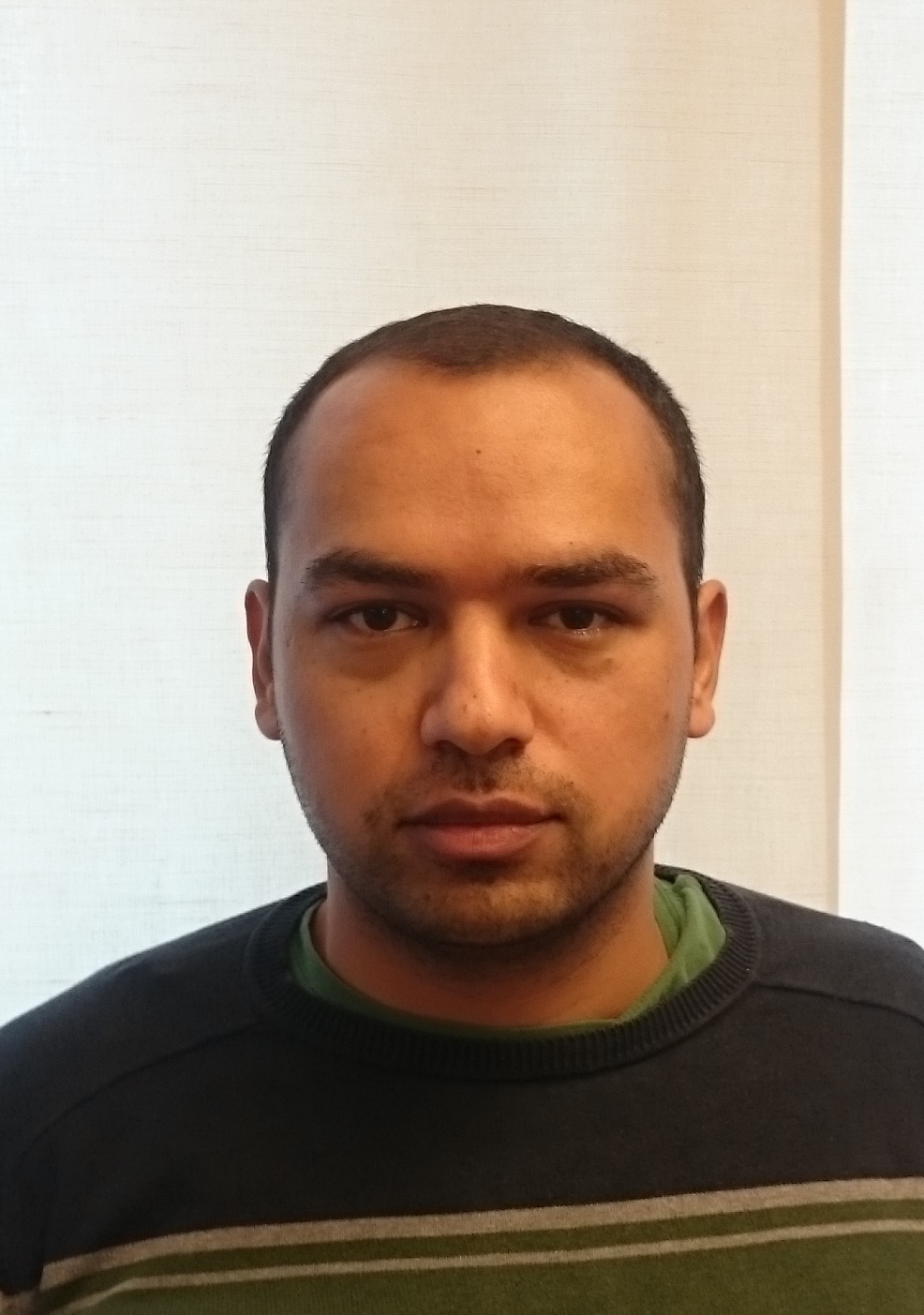}}]
{Ahmad A. Awad}
(Member IEEE) received his Ph.D. in Physics from the Autonomous University of Madrid, Spain, on high frequency magnetization and vortex dynamics in both magnetic and superconducting nanostructures in 2012. Since 2014, he has been a postdoctoral research fellow at the University of Gothenburg, Sweden.  His research interests include the characterization of magnetic nanostructures and spintronic devices with a focus on spin-torque induced magnetization dynamics. 
\end{IEEEbiography}

\begin{IEEEbiography}[{\includegraphics[width=1in,height=1.25in,clip,keepaspectratio]{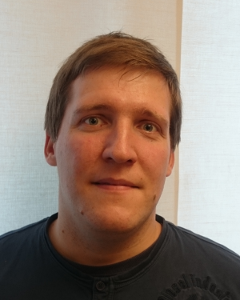}}]
{Philipp D\"urrenfeld}
(S'11) has a Dipl.-Ing. Univ. degree (2010) from the University of W\"urzburg, Germany. In 2015, he received his Ph.D. in physics from the University of Gothenburg, Sweden on the topic of spin-torque and spin Hall nano-oscillators with single magnetic layers. His research interests include the nanofabrication and electrical characterization of spintronics and magnonics devices.
\end{IEEEbiography}

\begin{IEEEbiography}[{\includegraphics[trim =  25mm 117mm 72mm 40mm width=1in,height=1.25in,clip,keepaspectratio]{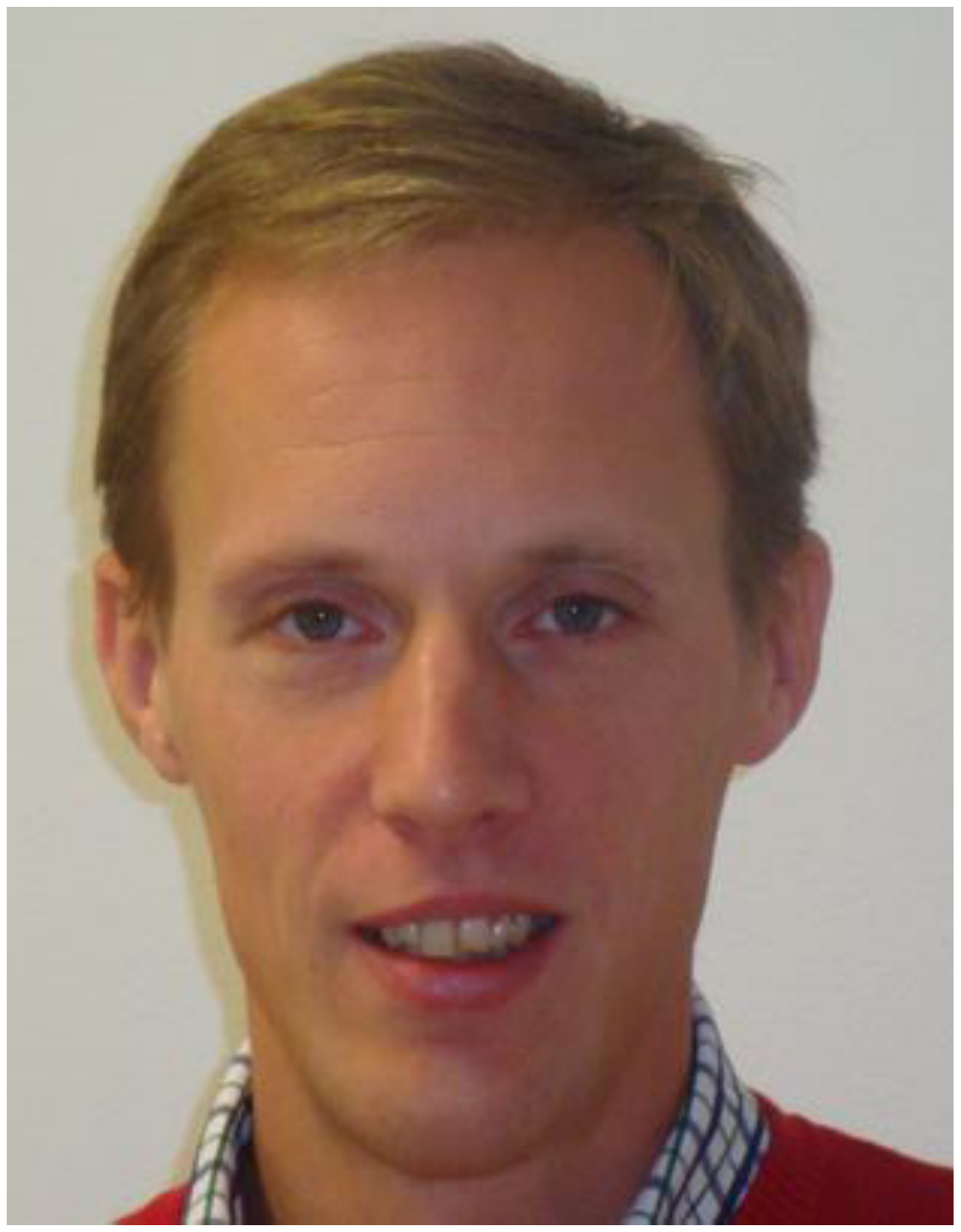}}]
{B. Gunnar Malm}
(M'98 - SM'10) was born in Stockholm, Sweden, in 1972. He received an M.S. from Uppsala University, Sweden in 1997  and a PhD in solid-state electronics from the Royal Institute of Technology (KTH), Stockholm in 2002. He has been an Associate Professor at the School of ICT, KTH since 2011. His recent work includes silicon photonics, silicon-carbide technology for extreme environments, and spintronics. He also serves on the KTH Sustainability Council.
\end{IEEEbiography}

\begin{IEEEbiography}[{\includegraphics[trim =  25mm 117mm 72mm 40mm width=1in,height=1.25in,clip,keepaspectratio]{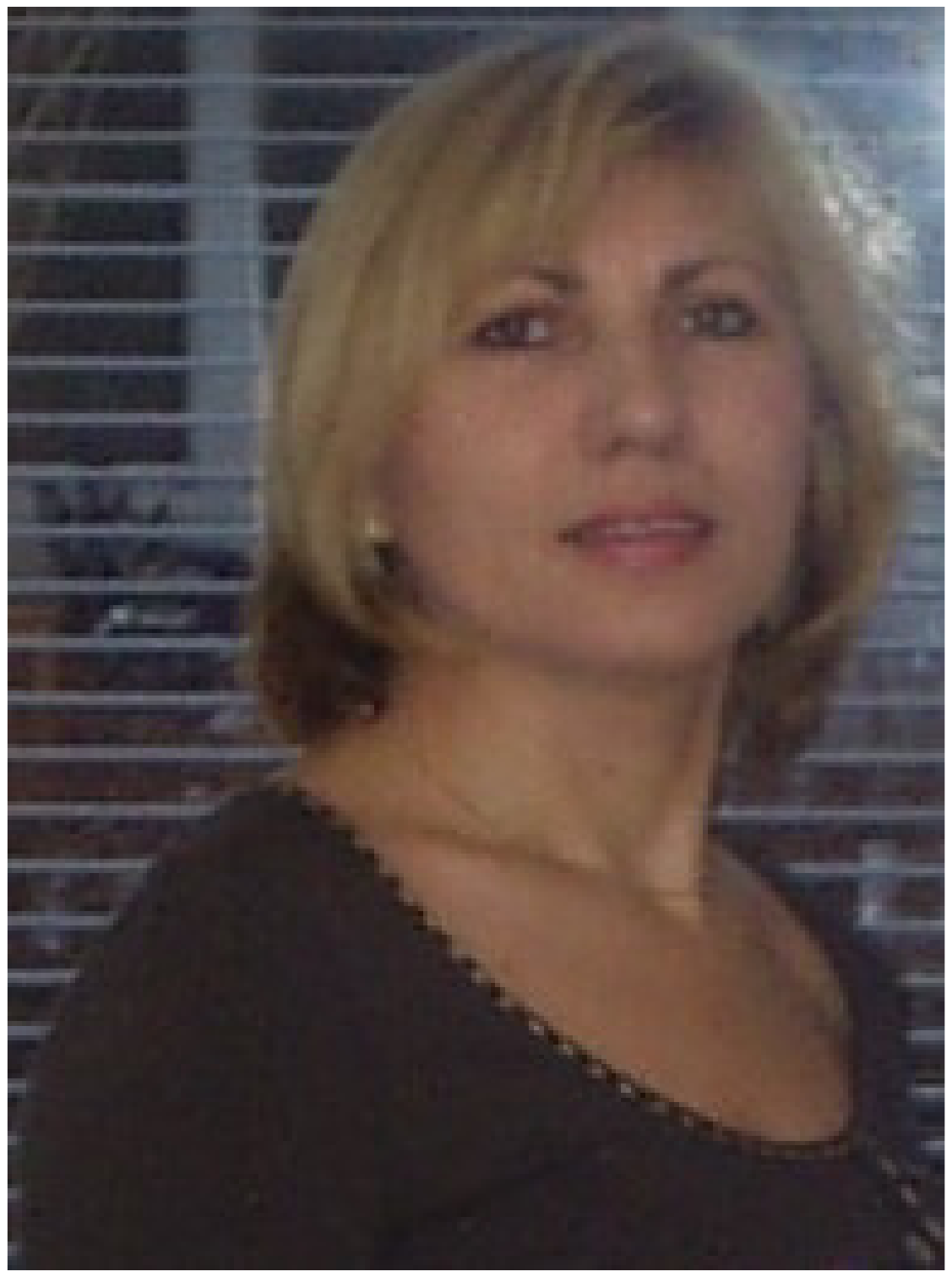}}]
{Ana Rusu}
(M'92) received an M.Sc. degree in electronics and telecommunications from the Technical University of Iasi, Romania in 1983, and a Ph.D. degree in electronics from the Technical University of Cluj-Napoca, Romania,in 1998.
She has been with KTH Royal Institute of Technology, Stockholm, Sweden, since 2001, where she is Professor in electronic circuits for integrated systems. Her research interests include low/ultralow-power high-performance CMOS circuits and systems, STO-based systems, RF graphene, and high temperature SiC circuits.
\end{IEEEbiography}

\begin{IEEEbiography}[{\includegraphics[trim =  25mm 117mm 72mm 40mm width=1in,height=1.25in,clip,keepaspectratio]{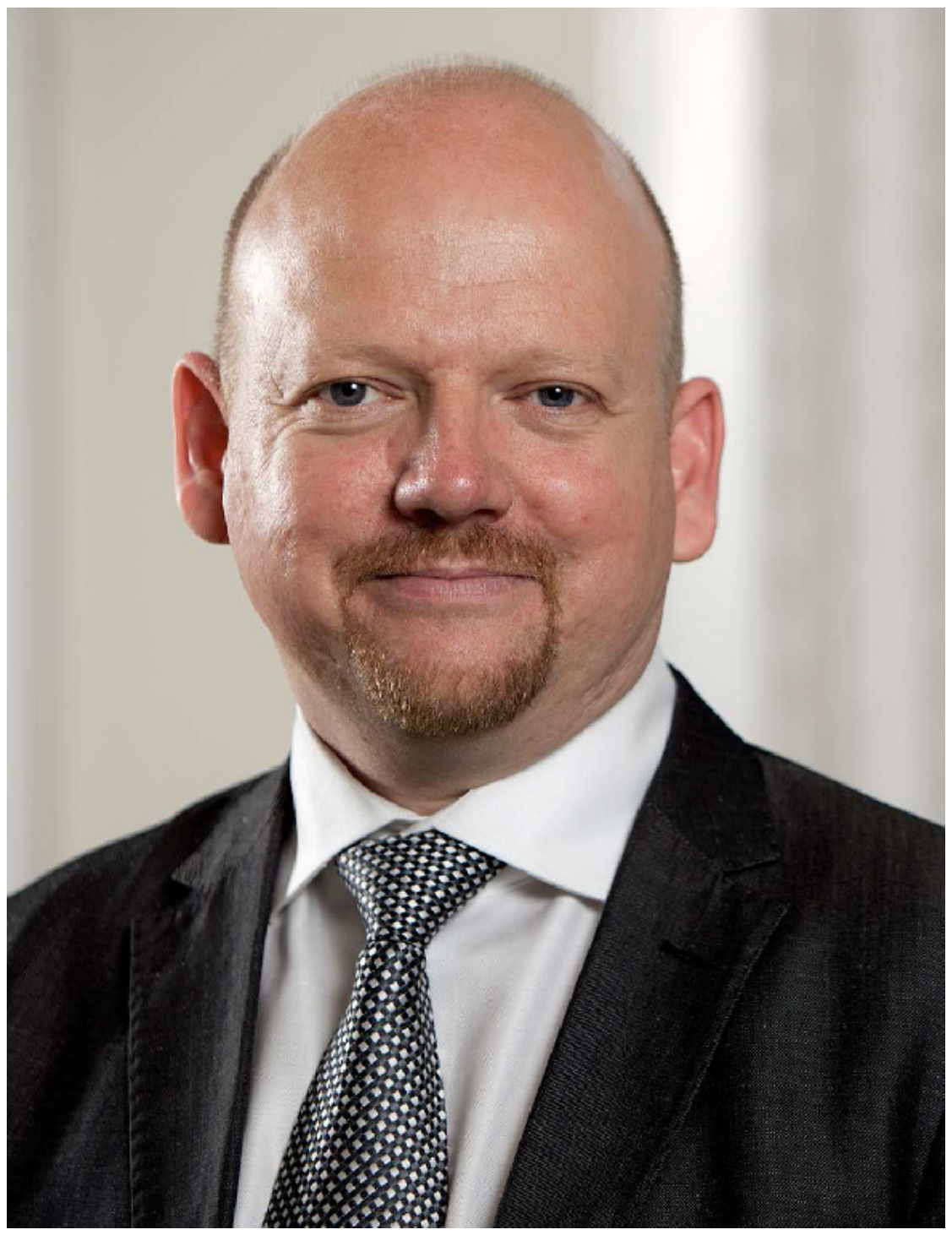}}]
{Johan $\AA$kerman}
(M'06) gained an Ing. Phys. Dipl. degree in 1994 from EPFL, Switzerland, an M.Sc. in physics in 1996 from LTH, Sweden, and a Ph.D. in materials physics in 2000 from KTH Royal Institute of Technology, Stockholm. In 2008, he was appointed Full Professor at the University of Gothenburg and is a Guest Professor at KTH Royal Institute of Technology. He is also the founder of NanOsc AB and NanOsc Instruments AB.
\end{IEEEbiography}


%\vfill


\end{document}